\documentclass[12pt,a4paper]{article}

\usepackage[british]{babel}

\usepackage[a4paper,top=2cm,bottom=2cm,left=2.5cm,right=2.5cm,marginparwidth=1.75cm]{geometry}

\usepackage[style=ieee, backend=biber]{biblatex}

\DeclareUnicodeCharacter{0308}{}

\usepackage[utf8]{inputenc}

\usepackage{amsmath}
\usepackage{graphicx}
\usepackage[colorlinks=true, allcolors=blue]{hyperref}
\usepackage{hyperref}
\usepackage[title]{appendix}
\usepackage{mathrsfs}
\usepackage{amsfonts}
\usepackage{booktabs} 
\usepackage{caption}  
\usepackage{threeparttable} 
\usepackage{algorithm}
\usepackage{algorithmicx}
\usepackage{algpseudocode}
\usepackage{listings}
\usepackage{enumitem}
\usepackage{chngcntr}
\usepackage{booktabs}
\usepackage{lipsum}
\usepackage{subcaption}
\usepackage{authblk}
\usepackage[T1]{fontenc}    
\usepackage{csquotes}       
\usepackage{diagbox}

\usepackage{setspace}
\usepackage{titlesec}
\titleformat{\section} 
  {\normalfont\Large\bfseries}{\thesection.}{1em}{}

\usepackage{float}   
\usepackage{caption} 


\title{Predicting the Emergence of the EV Industry: A Product Space Analysis Across Regions and Firms}
\author[1,2]{Katharina Ledebur}
\author[2,3]{Ladislav Bartuska}
\author[2,4]{Klaus Friesenbichler}
\author[2,1,5,*]{Peter Klimek}
\affil[1]{Complexity Science Hub Vienna, Metternichgasse 8, A-1030 Vienna, Austria}
\affil[2]{Supply Chain Intelligence Institute Austria, Metternichgasse 8, A-1030 Vienna, Austria}
\affil[3]{Logistikum, University of Applied Sciences Upper Austria, Wehrgrabengasse 1-3, A-4400 Steyr, Austria}
\affil[4]{Austrian Institute of Economic Research, Arsenal Objekt 20, A-1030 Vienna, Austria}
\affil[5]{Institute of the Science of Complex Systems, CeDAS, Medical University of Vienna, Spitalgasse 23, A-1090 Vienna, Austria}

\affil[*]{Corresponding author: \texttt{klimek@csh.ac.at}}
\date{}

\begin{document}
\maketitle
\begin{abstract}
The automotive industry is undergoing a profound transformation, driven by the electrification of powertrains, the rise of software-defined vehicles, and the adoption of circular economy concepts. These trends are increasingly blurring the boundaries between the automotive sector and other industries. 
The pace of adaptation to electrification varies considerably between regions and firms. Unlike internal combustion engine (ICE) production, where mechanical capabilities dominated, competitiveness in electric vehicle (EV) production increasingly depends on expertise in electronics, batteries, and software.
This study investigates whether and how firms’ ability to leverage cross-industry diversification contributes to their competitive advantage in this evolving landscape.
We develop a country-level product space covering all industries, and an industry-specific product space covering over 900 automotive components. This allows us to identify clusters of parts which are exported together, revealing shared manufacturing capabilities.
Closeness centrality in the country-level product space, rather than simple proximity, is a strong predictor of where new comparative advantages are likely to emerge. First, we examine this relationship across all industrial sectors to establish general patterns of path dependency, diversification and capability formation. Then, we focus specifically on the electric vehicle (EV) transition.
It is argued that new strengths in vehicles and aluminium products in the EU will generate 5 and 4.6 times more EV-specific strengths, respectively, than other EV-relevant sectors over the next decade. In contrast, these sectors are expected to generate only 1.6 and 4.5 new strengths, respectively, in already diversified China.
However, a different pattern emerges when these country-level results are compared to the firm-level product space. Countries such as South Korea, China, the US and Canada show the greatest potential for diversification into EV-related products. Established producers in the EU are likely to come under pressure.
These findings suggest that the success of the automotive transformation will depend on the ability of regions to mobilize existing industrial capabilities, particularly in related sectors such as machinery and electronic equipment.
\end{abstract}

\section{Introduction}
The automotive industry is at the centre of several transformation processes. First, there is a significant structural shift away from vehicles with fossil-fuelled internal combustion engines (ICEs) towards electric vehicles (EVs), in an effort to reduce greenhouse gas emissions \cite{kim2021role, maybury2022mathematical, dall2022low}. Second, there is a shift towards software-defined cars with the increasing importance of driver assistance systems, highly autonomous driving and in-vehicle infotainment systems \cite{sommer2021digital}. The electrification of the powertrain further strengthens the links between the automotive and electronical equipment industries through the increased demand for power electronical equipment in battery management systems \cite{burkacky2023outlook}. Thirdly, the recovery, recycling and reuse of end-of-life materials embedded not only in batteries but also in other automotive components is playing an increasing role with the shift towards circular value chains \cite{demartini2023transition}. Therefore, the process of structural change within the automotive industry cannot be properly assessed without taking into account the increasing integration with other industries. It is hard to overstate the importance of inter-industry collaboration and knowledge spillovers in the transformation process. In the fourth quarter of 2024, for example, the Chinese manufacturer BYD overtook Tesla to become the largest manufacturer of electric vehicles \cite{fu2024china}, despite having been founded 30 years earlier as a battery manufacturer \cite{wang2022specialised}. At the same time, established carmakers, particularly in the EU, are losing market share to new Chinese carmakers \cite{acea2024fact}. 

This uneven pace of change is accompanied by a shift in global demand. Worldwide, EV sales increased by 35 \% between 2022 and 2023, reaching around 14 million which translates to 18\% of all cars sold. However, almost 60 \% of new EV registrations in 2023 were in China. Europe and the United States accounted for just under 25 \% and 10 \% respectively \cite{global2024global}.  The reasons for this heterogeneous development include not only higher costs in the EU (in terms of labour and energy), but also a lagging technological capability \cite{draghi2024future}. 

The development of new technological capabilities can be understood in a quantitative and evolutionary way through path dependency and the principle of relatedness \cite{hidalgo2018principle, li2024evaluating}. This states that products that require similar technological capabilities, know-how and infrastructure tend to be co-exported by regions that have these capabilities \cite{Hidalgo2009}. It can be quantified in the product space, which is a network that quantifies the similarities of products based on the probability of their co-production or co-export by different countries~\cite{Hidalgo2007}. 
It provides a definition of economic complexity: regions with a large number of highly specialized skills can export a wide variety of products that few other regions can export \cite{hausmann2007you, Hidalgo2009, cristelli2015heterogeneous, mealy2019interpreting, sciarra2020reconciling}.
Such complex economies have the potential to diversify across large regions of the product space \cite{hausmann2014atlas}.
The idea is that regions are more likely to develop a strength in exporting certain goods if they already have this strength in goods that require similar capabilities, i.e., in products that are in close network proximity in the product space.
Recent research has shown that these and related indicators of complexity are proxies for hidden economic potential and therefore are predictive of economic growth \cite{Hidalgo2021, mewes2022technological}, although the predictive power has been found to vary across growth regimes \cite{cristelli2015heterogeneous}. Such indicators can be used to infer capabilities from the regional distribution of occupational categories, patents or skills \cite{mealy2019interpreting, Hidalgo2021}. Firm level evidence on capabilities mirrors such structural evidence and argues that knowledge-relatedness and path dependency are critical in firms’ technological diversification activities \cite{breschi2003knowledge}.

It is challenging to use complexity economics approaches to analyse industrial transformation processes in greater detail and policy relevance.
Studies using the whole product space typically draw on the Harmonised System classification of products \cite{hausmann2014atlas, cristelli2015heterogeneous}, which is useful at the macroeconomic level, but its limited resolution hampers the analysis of specific technologies or industries. 
The finest globally available resolution of the HS six digit data contains about 5,000 different products in the whole economy. There are estimations that a single car alone requires between 70,000 and 90,000 components\footnote{\url{https://www.spainautoparts.com/en/blog/how-many-components-does-a-car-have}, accessed 02/12/2025}.
On the other hand, more granular firm-level automotive databases offer a resolution of several hundred categories of components, and link these to the companies that produce them \cite{kito2014structure, Kito2018, brintrup2018predicting}. 
However, such industry-specific datasets do not inform about the relations between the car industry and other industries.

The aim of this study is to link industry-specific and economy-wide transformation studies. We use a component-based approach to quantify the competitive positioning and diversification potential of firms and regions in the transition from ICE- to EV-specific automotive components. Here, diversification potential refers to the capacity of a country, or firm, to develop new comparative advantages in related products based on existing industrial capabilities.

For this we employ two datasets. One on firm-level data on vehicle and component manufacturers worldwide, including detailed information on supplied parts, production networks, and technological specialization across more than 60,000 firms (Marklines data). The second dataset (BACI data), is developed by the CEPII research center and contains harmonized and highly detailed international trade data based on the UN Comtrade system, covering more than 5,000 products classified under the Harmonized System (HS) at the six-digit level.
We compare product spaces of high-resolution firm-level data for the automotive industry (Marklines data) and comprehensive country-level export data for all industries (BACI data).
In particular, we use a supplier database to construct a product space containing more than 900 components produced by more than  60,000 firms.
We map these components to their corresponding Harmonised System (HS2012) codes at the six-digit level, which represent standardized product categories used in international trade statistics. This mapping enables us to link firm-level component data to the global trade network, allowing the identification of related products and industries in the broader economy.

We use these product spaces to assess cross-industry automotive transformation pathways for different regions.
We classify car components according to whether they are used exclusively in EVs, ICEs or both technologies.
For each region, we can then assess how close its current industrial structure is to EV-specific components, both within the automotive product space and in other industries.
More specifically, we assess whether a product (component) is typically exported from a country (produced by a company in the country) and then measure its network distance to EV-specific components as a quantitative estimate of the diversification potential.
The quantitative framework is validated by assessing how well the principle of relatedness can be used to explain the past emergence of comparative strengths in specific products using historical export data in general and in the automotive industry in particular.
Furthermore, we assess how these transformation pathways could contribute to derisking by comparing these estimates with concentration indices of the corresponding products.

Previous research has used complexity-based approaches to study industrial transformation and diversification at national and sectoral levels\cite{yamada2024structure, bam2021io, cresti2025vulnerabilitiescapabilitieseuautomotive}. These studies have demonstrated that product spaces can capture technological relatedness and facilitate the identification of diversification opportunities within specific industries or countries. However, these studies are usually confined to national export structures or single-industry networks, which limits their ability to capture cross-industry dynamics. We address this issue by linking firm-level and economy-wide product spaces to identify diversification pathways within the automotive industry and across related sectors. Our contribution to the existing literature is twofold. First, we establish a link between industry-specific and economy-wide approaches in a product space context, making a methodological contribution. Second, our findings shed light on transition paths and network requirements for the development of a competitive electric vehicle (EV) industry. By identifying how existing industrial capabilities enable or constrain diversification into emerging technologies, our analysis provides insights into the economic resilience of regions, understood as their ability to adapt to structural change and maintain competitiveness. This is of practical relevance to firms and policymakers.

\section{Literature review}

\subsection{Complexity Economics}

The basic assumption of economic complexity theory (complexity economics) is that economic growth is to some extent driven by the capabilities (technology, know-how, capital, labor, etc.) of the region under consideration \cite{hausmann2007you}. 
The more capabilities a region has that are found in only a few other regions at the same time, the more complex it is. 
The degree of economic complexity implies a certain level of economic performance that regions will approach over time. 
In recent years, the theory of economic complexity has established itself as a standard method for many questions of economic development, but also for innovation studies \cite{Hidalgo2021}.

These mostly latent capabilities of a region are difficult to measure directly.
Hidalgo and Hausmann \textit{et al.}~\cite{Hidalgo2009} proposed a method to infer these capabilities from indicators of comparative strength using finely resolved export data.
They argue that there is a comparative strength in a product category if the share of exports in this category is higher than the corresponding share of total world exports \cite{balassa1963empirical}.

Interestingly, the export strengths of individual countries are shown to be 'nested' within each other~\cite{Hidalgo2009}. 
There are countries with highly diversified exports and comparative strengths across the product range. 
In contrast, there are less diversified regions whose strengths lie in products that are also among the export strengths of many other regions, while highly specialized products that are only exported by a few countries (e.g. microelectronics, pharmaceuticals, etc.) tend to be exported only by highly diversified regions.
The nested nature of comparative advantage arises because 'complex' products require many different skills and can therefore only be produced by complex economies, while 'simple' products requiring simpler skills can be produced in many places. 
Hidalgo and Hausmann proposed a set of complexity indicators, essentially measuring how many of these complex products a country exports, and showed that these indicators explain economic growth much more accurately than comparable indicators.
Since then, these ideas have been refined and extended to measure comparative advantages beyond exports, such as industrial sectors \cite{neffke2011regions}, labor flows \cite{neffke2013skill}, patents \cite{boschma2015relatedness}, digital products \cite{stojkoski2024estimating} and software \cite{juhasz2024software}.

\subsection{The Principle of Relatedness and the Product Space}

The theory of economic complexity argues that a region is more likely to develop comparative advantages in a given product if it already has strengths in products that require similar capabilities.
Which products these are can be determined by the principle of relatedness \cite{breschi2003knowledge, hidalgo2018principle, li2024evaluating}.
If two products require similar capabilities, it is expected that both will often be common comparative advantages of countries with the necessary capabilities.
Mathematically, this can be represented in a product space, which is a network in which nodes correspond to products and links connect products that are significantly more often comparative advantages of the same region than would be expected by pure chance \cite{Hidalgo2007}.

Formally, the product space is a one-mode projection of a specialization matrix, which in turn is constructed using a location-activity matrix.
That is, let $X_{cp}$ be the location-activity matrix indicating volume of activity (for instance, the exports of a product) $p$ in location $c$. 
A specialization matrix $R_{cp}$ is computed as
\begin{equation}
R_{cp} = \frac{X_{cp} X_{cp}}{X_c X_p}   \quad,
\end{equation}
where we used the shorthand notation that for any matrix $A_{ij}$, 
$A_i \equiv \sum_j A_{ij}$.
$R_{cp}$ quantifies the ratio between the observed and expected level of economic activity. 
A binary specialization matrix $M_{cp}$ is defined as
\begin{equation}
M_{cp} =
\begin{cases} 
1, & \text{if } R_{cp} \geq 1 \\
0, & \text{otherwise}
\end{cases}
\end{equation}
This transformation retains only those location-activity relationships that exceed the expected specialization threshold. The product space is then constructed as follows ~\cite{Hidalgo2021},
\begin{equation}
    \Phi_{pp'} = \frac{\sum_c M_{cp}M_{cp'}}{max(M_p, M_p')} \quad.
\end{equation}

The product space provides an intuitive framework for quantitatively assessing regional diversification potential.
Relatedness suggests that regions typically develop comparative advantages in new products that are close to the products in which they already have strengths (and therefore require similar capabilities). 
This is reflected by the product space, which has a clear core-periphery structure. 
Complex products are exported together with many other products and form a tightly connected core in the network. 
Simple products, which require only a few capabilities, enable only a few other products and are therefore located at the periphery of the network with significantly fewer connections. It follows that a region with more complex products also has access to more potential future export strengths than a region whose products are exclusively located at the periphery of the product space. 
Indeed, complexity indicators can also be interpreted as spectral clustering methods that indicate whether a country's products are located in the densely connected core or in the periphery of the product space \cite{mealy2019interpreting}.

\subsection{Green Complexity}

The complexity economics approach can also be extended to specific subsectors of the economy. 
For example, product categories that are considered key technologies for the green transition can be weighted more heavily to create a 'green complexity indicator'~\cite{mealygreen}. 
This indicator shows the extent to which such key technologies are already being exported in a region.  
The potential of a region to further diversify in such green technologies is measured by the 'Green Complexity Potential' ~\cite{mealygreen}. There is a strong correlation between a country's existing green complexity and its potential for future green diversification, which in turn implies a strong path dependency in the development of related capabilities~\cite{Andres2023}.

\subsection{Complexity Economic approaches to the automotive industry}

While the theory of complexity economics was originally developed to study problems in development economics, its scope has recently increased to sectoral industrial policy.
Yamada et al. constructed a product space by measuring firm-level product relatedness within a dataset of 11 Japanese carmakers and their suppliers \cite{yamada2024structure}.
They show that the resulting product space shares several topological characteristics with international product spaces derived from export data. These include a core-periphery structure, the emergence of new products in close network proximity to pre-existing comparative advantages, and the presence of capabilities conducive to diversification into new products.
The analysis therefore suggests a fractal structure of product spaces, in which the development of economies can be described by similar means on a domestic, firm-based and on a more aggregated, international level.

Bam et al. followed a different approach by analyzing the South African automotive value chain by means of an input--output (IO) product space \cite{bam2021io}.
In brief, this approach measures comparative advantages from export data and then integrates the results with a value chain perspective.
Therefore, the product codes are mapped to their industries, which enables one to measure the economic complexity not only of specific industries, but also of its up- and downstream sectors.
Bam et al. used this approach to identify specific sub-sectors within the automotive industry that are likely to support both short- and long-term economic growth.

Cresti et al. also identify product codes related to the automotive industry and analyse the economic complexity rankings of EU countries for products related to ICE, EV and hybrid vehicles \cite{cresti2025vulnerabilitiescapabilitieseuautomotive}.
They find high diversification potential in the automotive supply chain for countries such as the Czech Republic, Italy, Poland, France and Germany.
They further complement the analysis with IO data to measure dependencies on other sectors and with concentration indices derived from trade data to assess vulnerability.
They find a high degree of vulnerability combined with a relatively low potential for diversification in electrical accumulators for the EU.

The analyses by Yamada et al. on one side and Bam et al. and Cresti et al. on the other side have to some extent complementary strengths and weaknesses.
Approaches that are purely based on trade data or, even higher aggregated, sector-level information, offer a limited resolution of individual industries.
Such product spaces are often constructed at the four-digit level of product categories, where there is one code for cars and one for car parts. 
On the six-digit level (the most granular level offered by publicly available global trade datasets) these categories can only be disaggregated into a handful of subcategories.
This limitation can be overcome using firm-level data.
However, such datasets often cover only the automotive industry.
Hence, they do not allow to assess links with other sectors and cannot be used to study cross-industry structural change.

In the current work we develop a new framework that combines these two approaches.
We construct a high-resolution firm-level product space for the automotive industry which is linked to an economy-wide product space and therefore to other industries.

\section{Methodology}
\subsection{Data and industry- and firm-level product spaces}

The BACI dataset builds on the trade data from the United Nations Statistics Division Comtrade dataset and reports per country from 2007 to 2023 export values on six-digit level Harmonized System (HS) codes.
The industry-level export data product space $\Phi^{X}$ is computed from this data using a location-activity matrix $X_{cp}$ where each entry describes the export volume of product $p$ from country $c$ in 2022.

Our study also uses data on suppliers of automotive products from Marklines, a widely used online platform.\footnote{\url{https://www.marklines.com/}, accessed 3/3/2025} This dataset provides information on 62,198 companies in 86 countries, covering 921 different products (components). The database uses a hierarchical categorisation system for components. For example, connecting rods are part of the engine structure category, which is part of the broader engine category. In this study we work with the third tier of this hierarchy to ensure fine-grained results. These components are categorized in EV-specific, ICE-specific and unspecific components. Additionally we assign standardized product codes from the Harmonised System (HS) coding.

The firm-level product space $\Phi^{FL}$ is constructed from a binary location-activity matrix, $X_{fp}$, where each entry indicates if a specific firm $f$ produces product $p$. Using the corresponding specialization matrix $M_{fp}$ the product space $\Phi^{FL}$ is
\begin{equation}
    \Phi^{FL} = \frac{\sum_f M_{fp}M_{fp'}}{max(M_p, M_p')} \quad.
\end{equation}

\subsection{The Automotive Product Space}

\begin{figure}
    \centering
    \includegraphics[width=.99\linewidth]{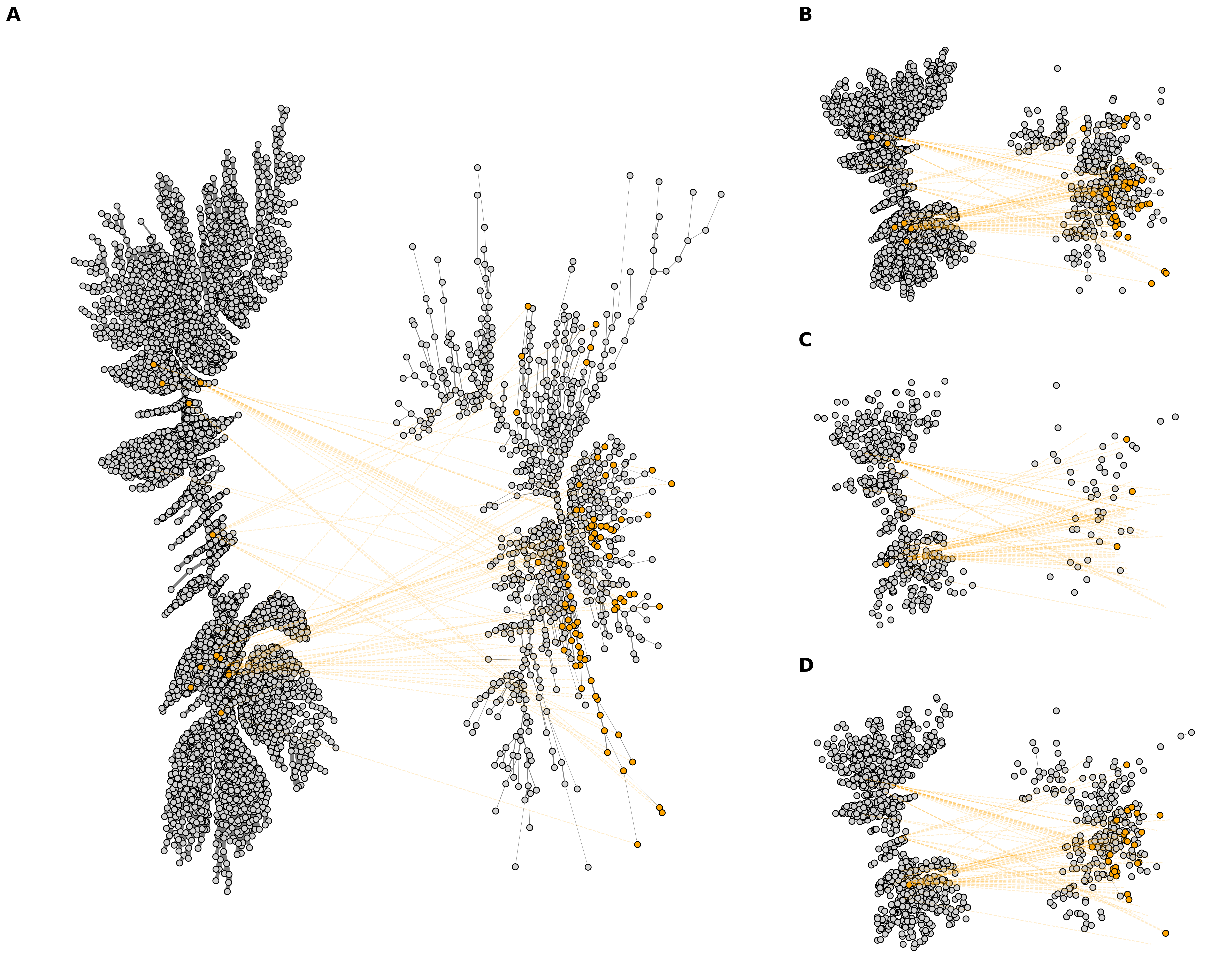}
    \caption{Product Space of industry- ($\Phi_X$) and firm-level ($\Phi_FL$) data . Shown are (A) the complete product spaces, and the country-specific product spaces for (B) China, (C) New Zealand and (D) Hungary. These countries are in the 100th, 65th, and 40th percentile, respectively. The network on the left hand side stems from the export data product space ($\Phi^X$) and the n etwork on the left hand side stems from the firm-level data $\Phi^{FL}$ is shown in blue. EV-specific nodes are colored in orange. Dashed orange links show the EV-specific nodes in $\Phi_FL$ mapped to EV-specific nodes in $\Phi_X$.}
    \label{fig:bacimlnw}
\end{figure}

The structure of the automotive firm- and industry-level product space is shown in Figure~\ref{fig:bacimlnw}(A).
The network on the left shows nodes in the trade data layer are linked by co-exports ($\Phi^X$) and the network on the right, is the firm-level layer ($\Phi^{FL}$). 
EV-specific products from the trade layer that are connected via inter-layer links to components in the firm-level layer are highlighted in orange.
These product spaces enable the mathematical representation and visualisation of a country's industry structure.

To this end, a product space from the industry-level product space $\phi_{X}$ is constructed for each individual country or region by "cutting out" only those nodes that correspond to its comparative advantages and also dropping all of their links.
In network-theoretic terms, we construct for each country the induced subgraph of the product space containing all nodes that are comparative advantages.
Examples of the resulting country-specific networks are shown for (B) China, (C) New Zealand, and (D) Hungary, where nodes not corresponding to strengths are grayed out. These countries are in the 100th, 40th, and 65th percentile, respectively.
It is evident from an initial observation that China and Tunisia differ in terms of their diversity and the number of comparative advantages.

Figure~\ref{fig:ml_productspace} shows a network visualization of the automotive product space constructed from the firm-level dataset, $\Phi^{FL}$.
Nodes correspond to components with colors indicating whether they are EV-specific (orange), ICE-specific (red) or unspecific (grey).
Links indicate that two components tend to be manufactured by the same companies.
Out of the 917 components comprising the product space, 78 are EV- and 265 are ICE-specific.
Both of these types form clearly visible clusters in the network, suggesting that they indeed require different skills.

\subsection{Operationalizing opportunity and derisking into indicators}

The revealed comparative advantage (RCA) is the ratio of a country's share of world exports of a given product to that product's share of world exports of all products. It indicates a country's competitiveness and specialisation in a particular product. The standardized RCA (sRCA) \cite{laursen2015revealed} transforms the RCA into values ranging between -1 and 1. \\
We build on the Green Complexity Potential (GCP) framework introduced by Mealy and Teytelboym~\cite{mealygreen}, adapting it to the context of the automotive industry. While the original GCP aims to identify countries with the potential to competitively produce or export "green" products, our focus is on three distinct product categories: EV-specific products, ICE-specific products, and products used in both technologies (referred to as unspecific products).

To this end, we develop three corresponding indicators: EV-Complexity Potential (EVCP), ICE-Complexity Potential (ICECP), and Unspecific-Complexity Potential (UCP). These indicators measure how closely a country's existing production structure aligns with the technological requirements of electric, combustion, or non-specific automotive products. Based on current industrial capabilities, they provide a forward-looking measure of the potential to diversify into emerging technologies. In the following, we outline the methodology for computing EVCP; the same approach is applied analogously to the other two categories.

For the firm-level dataset, we calculate EVCP at the firm level (based on the specific components each firm manufactures) and then aggregate these values by taking the country average. For the industry-level dataset, we compute the EVCP directly at the country level. In both cases, we standardize the resulting EVCP scores such that values above (below) zero indicate higher (lower) than average potential.

The EVCP measures the potential for a country to diversify into green or EV-specific products it is not currently competitive in, based on the proximity of these products to the country's existing production or export basket and the Product Complexity Index (PCI), which reflects the knowledge intensity of each product, 
\begin{equation}
    \text{EVCP}_c = \frac{1}{\sum_{g \notin G_c}(1-\rho_g^c)} \sum_{g \notin G_c} (1-\rho_g^c) \phi_{gc} \tilde{\text{PCI}}_g
\end{equation}
where \( g \notin G_c \) are EV-specific products \( g \) in which country \( c \) is not competitive (\(\text{RCA} \leq 1\)) in and \( \phi_{gc} \) is the proximity of product \( g \) to country \( c \)’s current export basket,
\begin{equation}
    \phi_{gc} = \frac{\sum_{i \in G_c} \phi_{gi}}{|G_c|}
\end{equation}
with \( \phi_{gi} \) as the conditional probability of exporting \( g \) given \( i \), defined as
\begin{equation}
    \phi_{gi} = \min \left( P(g|i), P(i|g) \right) \quad.
\end{equation} \\

We further use the closeness centrality of all products specific to a country (RCA > 1) to EV-/Un-/ICE-specific products to quantify cross-industry diversification pathways. For each country we compute a specific product subspace which contains only links if at least one of the products connected by the link is a comparative advantage of the country. We define a node as reachable from a given node if there is a directed path on the network connecting the nodes. 

The closeness centrality ($C_c$) of country $c$ then measures how close a node, for instance an EV-specific product, is to all other nodes (products) in the industry-specific product subspace,
\begin{equation}
    C_c(i) = \frac{N-1}{\sum_{j \in R_c(i)} d_c(i,j)} \quad,
    \label{eq:close}
\end{equation}
where $N$ is the total number of products (nodes) in the network, $R_c(i)$ is the set of nodes reachable from product $i$ for country $c$ and $d_c(i,j)$ is the inverse proximity (distance) between product $i$ and $j$ in the corresponding product subspace. 

We operationalize the diversification potential through network-based measures of relatedness, viz closeness centrality and complexity potential, which capture how existing capabilities enable entry into new products.

To compare predictive performances of potential metrics, we estimate a logistic regression model with the outcome being the emergence of a new comparative advantage in a given product.

For each product $p$ and country $c$, we therefore introduce a variable $y_{c,p}$ that indicates whether a switch from a sRCA $\leq$ 0 in 2012 to sRCA > 0 in 2023, has taken place.

We evaluate the performance of three different metrics designed to capture proximity or closeness in the product space:
\begin{itemize}
    \item Product-level closeness centrality $C_p$. This metric measures how close product $p$ is to other products in the product space, based on 2012 data. Specifically, it is defined as the average inverse distance from $p$ to all products that were already comparative advantages of country $c$ in that year.
    \item Product-level complexity potential $CP_p$. This metric estimates the potential of country $c$ to develop a comparative advantage in product $p$ in 2012. It is calculated as the product of three components: (i) the average proximity of $p$ to products that $c$ already exports competitively, (ii) an indicator that $p$ is not already exported by $c$, and (iii) the normalized complexity of $p$.
    \item Product-level potential $P_p$. Similar to $CP_p$, this metric also measures the potential of country $c$ to export product $p$ competitively in 2012. However, it excludes product complexity from the calculation. It is defined as the product of (i) the average proximity of $p$ to products that $c$ already exports competitively and (ii) an indicator that $p$ is not yet exported by $c$.
\end{itemize}

We estimate a model for each of these explanatory variables and additionally control for (i) the product's export value in 2012 (log-scale) to account for initial activity levels and (ii) the number of unique products the country competitively exports.

We seek to identify the metric that is best suited to explain the emergence of competitive strengths in any of the twelve EV-specific HS codes. Therefore, we first estimate these models separately for each HS chapter, using sector-specific sRCAs that are calculated based only on products within the respective chapter. For each chapter and model specification, we conduct three sets of analyses: (1) using twelve randomly selected products from the chapter, (2) using twelve randomly selected products from those in the top quartile of Product Complexity Index (PCI), and (3) using twelve randomly selected products from the bottom quartile of PCI. If fewer than twelve products are available in a given quartile, we include all available products in that group.
In the next step we then estimate these models for the twelve EV-specific products only.

\subsection{Estimating the expected number of new comparative advantages}

Our framework allows us to estimate the number of newly emerging comparative advantages in a sector (such as EV products), given strengths in another sector.
Therefore, we first compute the closeness centrality between products in each HS chapter and EV-specific products. For each country \( c \) and HS chapter \( h \), we calculate: 

\begin{itemize}
    \item \( C_{c,h} \): the closeness between products in chapter \( h \) and EV-specific products,
    \item \( \bar{C}_c \): the average closeness across all chapters for country \( c \) and EV-specific products.
\end{itemize}
The difference $\Delta C = \bar{C}_{c,h} - \bar{C}_c$ measures whether products from sector $h$ are closer ( $\Delta C>0$) or further apart ( $\Delta C<0$) from EV products than the average over all chapters.
From this one can estimate the probability of developing new export strengths based on the logistic regression model. More concrete, we use the estimated logistic regression coefficient \( \beta \), the intercept term, and the standard deviation \( \sigma_C \) of closeness used in the model. The standardized change in closeness to EV products for country $c$ if it develops a strength in sector $h$ is then,
\[
X_{\text{std}} = \frac{\bar{C}_{c,h} - \bar{C}_c}{\sigma_C}
\]

The resulting change in the probability for the emergence of a new EV product follows as

\[
\Delta P = \text{sigmoid}(\beta X_{\text{std}} + \text{intercept}) - \text{sigmoid}(\text{intercept}) \quad.
\]

Finally, $\Delta P$ is multiplied by the average number of EV-specific products not yet competitively exported by the country in 2012, $\bar{N}_{\text{notheld}}$, to obtain an estimate of the expected number of newly enabled export strengths,

\[
n_y = \Delta P \cdot \bar{N}_{\text{notheld}} \quad.
\]

We compare the resulting number of new EV-specific strengths when gaining strengths in a given sector to other EV-relevant sectors over the next ten years. We define EV-relevant sectors as vehicles (HS chapter 78), electrical equipment (HS chapter 85), machinery (HS chapter 84), iron/steel articles (HS chapter 73), iron \& steel (HS chapter 72), aluminum products (HS chapter 76), copper products (HS chapter 74), organic chemicals (HS chapter 29), inorganic chemicals (HS chapter 28), misc. chemicals (HS chapter 38), plastics (HS chapter 39), rubber (HS chapter 40), optical \& medical instruments (HS chapter 90), aircraft \& spacecraft (HS chapter 88), jewels \& coins (HS chapter 71), paper \& articles (HS chapter 48), furniture \& lighting (HS chapter 94), and glassware (HS chapter 70).
\subsection{Import concentration}

The Herfindahl-Hirschman Index (HHI) is calculated to measure the import concentration for each country $c$ and EV-specific HS code $h$ as $HHI_{c,h} = \sum_{i}(v_{i,c,h}/V_{c,h})^2$, where $v_{i,c,h}$ is the import value from country $i$ and $V_{c,h}$ is the total import value. Additionally, a single HHI is computed for the EU. We show HHI values normalized by the global average HHI value.

\section{Results}
This section presents the main findings. First, we examine how the global automotive industry has shifted towards EV production by analyzing export data. Next, we assess whether a country's current industrial structure can explain the development of new comparative advantages and identify where diversification into EV-related sectors is most likely to occur. Finally, we assess how these opportunities relate to the risks of concentrated import dependence.

\subsection{The structural transformation of the automotive industry in trade data}

\begin{figure}[H]
    \centering
    \includegraphics[width=.99\linewidth]{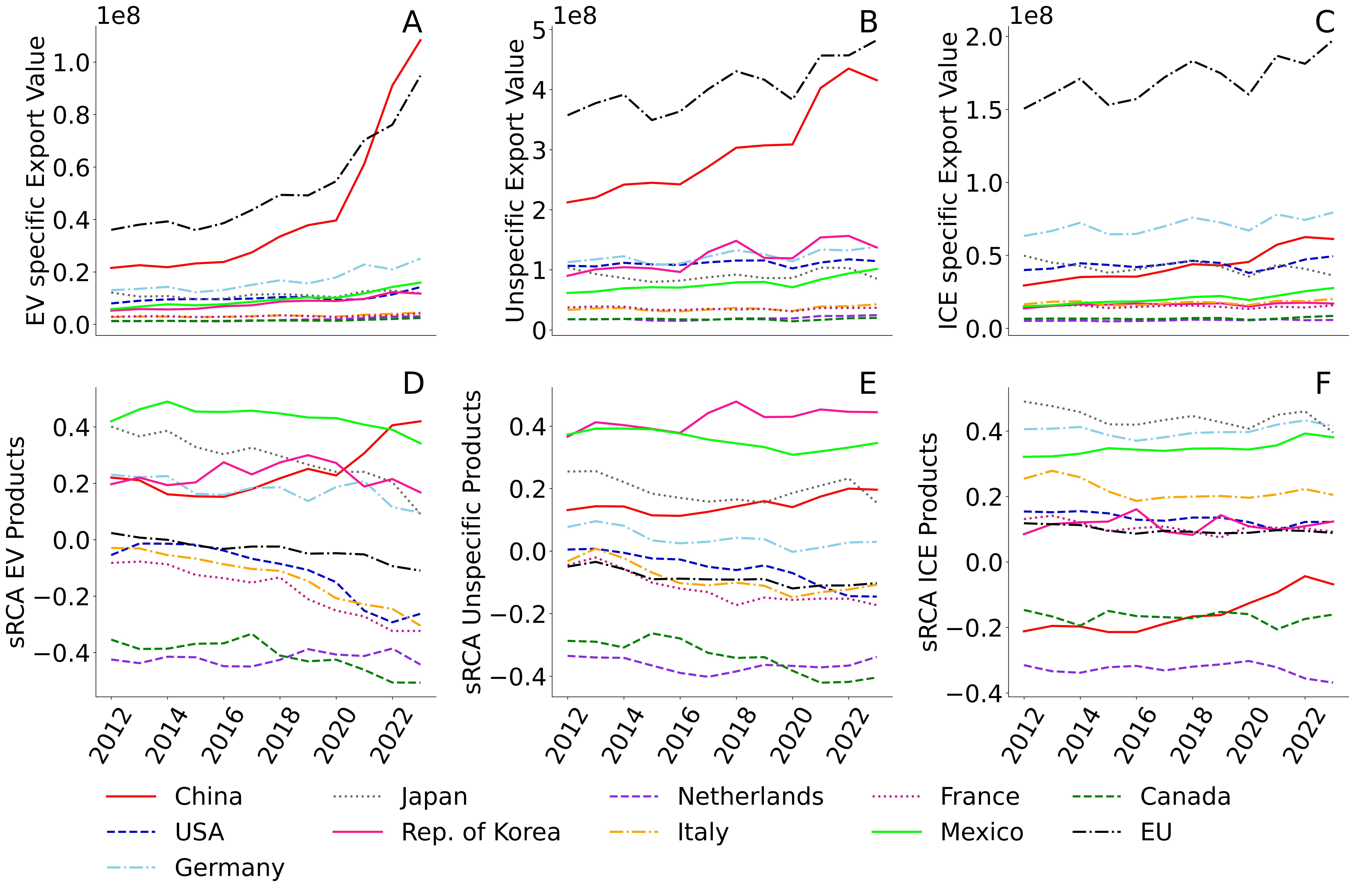}
    \caption{Trends in the export data of car components. Export value of EV specific HS codes (A), unspecific HS codes (B) and ICE specific HS codes (C), sRCA values for EV specific HS codes (D), unspecific HS codes (E), and ICE specific HS codes (F)
    for the 10 countries/regions with the highest EVCP in 2022. 
    In (D)-(F) values for the EU are shown by averaging over all EU countries weighted by each country's share of total EU exports in 2022.
    }
    \label{fig:bacircaexport}
\end{figure}

The structural transformation of the automotive industry with respect to EV adoption is clearly visible in the trade data.
Figure \ref{fig:bacircaexport} shows the combined export value of all (A) EV-specific, (B) unspecific and (C) ICE-specific products over time for selected countries and regions (the ten largest exporters and the EU as an aggregate are shown).
It is evident that Chinese exports exhibit a "hockey stick" trend in EV-specific car components, with a pronounced increase in recent years, surpassing the corresponding EU exports. This pattern is less pronounced, yet evident, in EU and Chinese exports for unspecific and ICE-specific products. Most time series show comparable stable developments over time. China and the EU exhibit the highest total export values.

A range of heterogeneous developments can be observed in terms of comparative advantages, as quantified by sRCA values.
The hockey stick phenomenon is also evident in Chinese EV products, see Figure \ref{fig:bacircaexport}(D).
Mexico, Japan, Korea, and Germany also demonstrate comparative advantages in EV-specific products, although there is a downward trend in recent years.
Most other countries show negative and decreasing RCA values.
For the EU, the indicators have shifted from comparative advantage to comparative disadvantage.

China is has been increasing its comparative advantages in unspecific products, Figure \ref{fig:bacircaexport}(E), a trend also exhibited by Korea. Mexico and Japan also consistently show comparative advantage.
The EU's sRCA values are negative and have further decreased over time.
China, while still negative, has rapidly been increasing its comparative advantages in ICE-specific products, Figure \ref{fig:bacircaexport}(F). 
In this regard, Japan, Germany, Mexico and Italy show high and stable values.

\subsection{Validating the explanatory power of the complexity indicators}

To validate our methodology, we quantify whether comparative advantages in one region of the product space explain the emergence of new strengths in close network proximity. 
Hence, we implement a logistic regression model per HS chapter. The outcome is the emergence of a new comparative advantage in a given product of that chapter from 2012 to 2022. All specifications control for the log of a product's export value in 2012 to account for initial activity levels, and for a country's export diversity, defined as the number of unique products it exports competitively.

The first explanatory variable we consider is a product's closeness to product $p$ in the year 2012 ($C_p$), measured as the average inverse distance to products in the product space that were comparative advantages of country $c$ in that year. 
The second and third explanatory variables are the product-level complexity potential $CP_p$ and the product-level potential $P_p$, respectively.
Both of these indicators only consider a product's direct neighbors in the product space; the complexity potential puts a higher weight on complex products.

We estimate the association for each HS chapter and explanatory variable using three sampling strategies to randomly select twelve products (or all available products if fewer than twelve are available) from: (1) the entire chapter; (2) the top PCI quartile; and (3) the bottom quartile. The regression results indicate that closeness ($C_p$) is the most consistent and significant predictor across all three sampling strategies (Tables~\ref{tab:regr_random}, \ref{tab:regr_toppci}, and \ref{tab:regr_botpci}). 
Across sampling strategies, $C_p$ yields the highest share of significant associations. For example, for fats \& oils (HS chapter 15), perfumes \& cosmetics (33), glassware (70), jewels \& coins (71), and vehicles (87), $C_p$ consistently shows positive and significant relationships with export performance. This suggests that structural embeddedness in the existing product space plays a central role in the development of competitive products. While $C_p$ remains significant in many chapters regardless of the sampling strategy, results from the top PCI sample tend to yield larger coefficients and more frequent significance for all three predictors. In contrast, the bottom PCI sample shows fewer significant results overall, suggesting that product space proximity is less important in the development of new export strengths in low complexity products.

\begin{figure}
    \centering
    \includegraphics[width=0.9\linewidth]{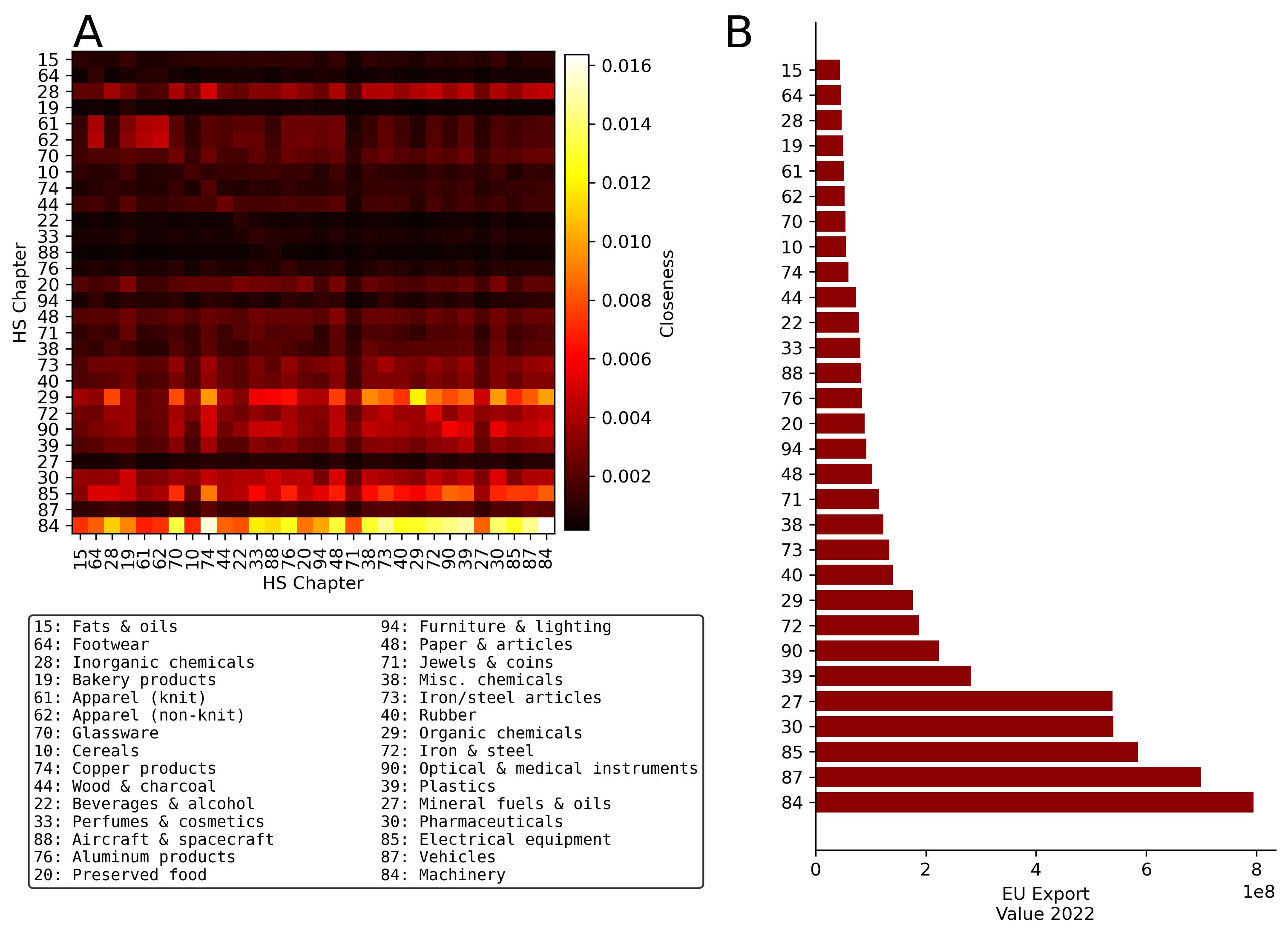}
    \caption{EU Closeness. A) Normalized closeness between the 30 HS chapters with the highest EU export value in 2022, measured as the average pairwise closeness between products in each chapter pair (i.e., sum over $C_{ij}$ normalized by $N_j$.) B) 30 HS chapters with the highest EU export value in 2022.}
    \label{fig:eu_close}
\end{figure}

Figure~\ref{fig:eu_close}(A) shows the average closeness between each pair of product groups (first two digits of the HS codes) in the EU, sorted by their total export value, all values taken in 2022 (B).
Three sectors stand out in terms of their closeness to other industries: machinery (HS codes 84), electrical equipment(85) and organic chemicals (29). 
These sectors show high closeness centralities with respect to all other sectors, which mirrors their central position in the EU product space.

Closeness centrality in other regions can produce different results.
Figure~\ref{fig:closeness_baci_descriptive} shows the ratios of sector-specific centralities in the EU compared to the USA and China.
Compared to the USA, the EU has much higher closeness centrality scores for most sectors, including perfumes \& cosmetics, aircraft \& spacecraft, jewels \& coins, chemicals, optical \& medical instruments, plastics, and pharmaceuticals, while lower scores are seen for footwear and apparel.
Compared to China, EU industries are closer to footwear, apparel, glassware, furniture \& lighting, iron \& steel articles, as well as electrical equipment.

Overall, these findings highlight the strong path dependence of industrial upgrading and the central role of proximity in shaping countries' ability to enter EV-related markets.

\subsection{Assessing opportunities in the automotive industry}

When we only use the emergence of comparative advantages in EV products rather the entire product range, we obtain qualitatively similar results.
The regression indicates a positive and statistically significant association between EV specific closeness and switch to sRCA > 0 from 2012 to 2023 ($\beta$ = 0.418, p = 0.0004), see Table ~\ref{tab:regr_ev}. This coefficient implies that a unit standard deviation increase in closeness to EV-specific products in 2012 increases the odds of developing comparative advantages in EV products by 52\% . 

\begin{figure}
    \centering
    \includegraphics[width=1\linewidth]{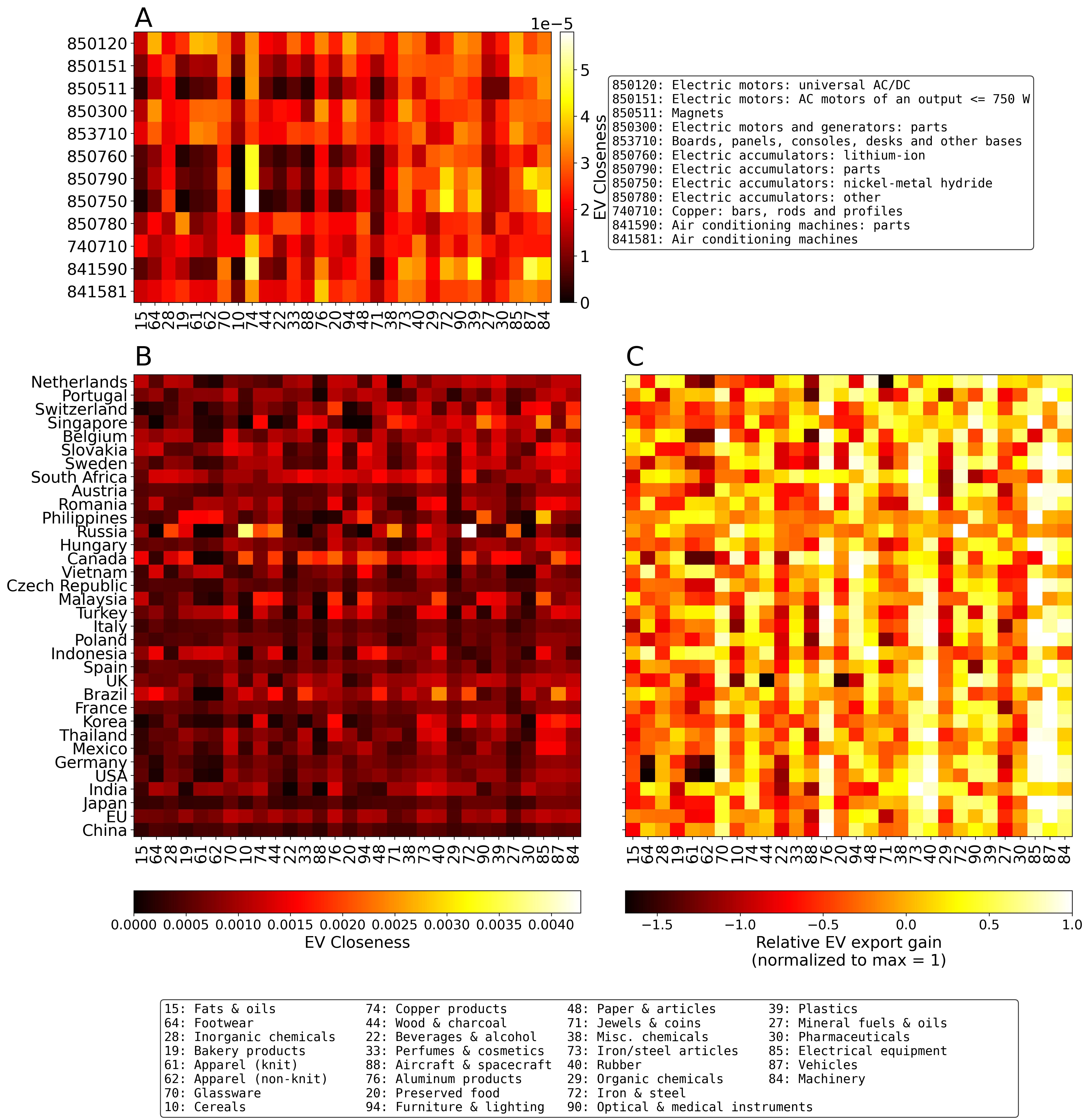}
    \caption{Closeness gain to EV specific products. A) Normalized closeness between the 30 HS chapters with the highest EU export value in 2022 to EV specific products, measured as the average pairwise closeness between products in each chapter pair (i.e., sum over $C_{ij}$ normalized by $N_j$.) B) Average closeness to EV-specific products among the top 25\% of products (by closeness) per country and HS chapter for the 30 HS chapters with the highest EU export value in 2022. C) Expected new EV strengths 30 HS chapters with the highest EU export value in 2022 normalized to maximum export gain per chapter and country.}
    \label{fig:gain_ev}
\end{figure}

Figure~\ref{fig:gain_ev}(A) shows how close the EU is to EV-specific products in each industry (HS chapter).
There are two main types of relationship. For permanent magnets (HS code 850511) and EV batteries (HS codes with 8507), only a limited number of other closely related sectors exist, including machinery and vehicles, as well as different materials and related products (e.g., plastic and copper).
In contrast, for other EV products, closeness centrality tends to be much higher across all chapters, suggesting that these products are easier targets for diversification.

We compute the average closeness centrality ($C_{c,h}$) to EV specific products per HS chapter ($h$) and country ($c$), see Figure~\ref{fig:gain_ev}(B). We averaged the products per chapter that were in the top 25\% in terms of closeness. See the supplementary information for the results using the top 10\% and 50\%.
Considering the existing comparative advantages of these countries, the machinery, vehicles, electrical equipment, rubber products, and iron/steel \& articles sectors emerge as key players in EV value chain diversification.

Combined with  the increase in probability of gaining another strength in any other product in the next ten years, $\Delta P$ , resulting from the regression model, and the average number of products in this chapter a country is not yet specified in, $\bar{N}_{notheld}$, we can estimate the number of expected new EV strengths $n_y$ over the next 10y for each newly acquired strength per country and HS chapter. We show the results relative to the maximum number of strengths per country Figure~\ref{fig:gain_ev}(C).  

Evaluating these results for specific sectors in the EU, we find that acquiring a new comparative advantage in vehicles (HS chapter 78), aluminum products (HS chapter 76), and rubber (HS chapter 40) yield the highest diversification potential. For instance if taking our regression results as causal, acquiring a new comparative advantage in vehicles leads to 5 times more EV strengths compared to the average EV-relevant sector. A new comparative advantage in aluminum or rubber products would lead to 4.6 and 3.7 times more new EV-specific strengths, respectively. 

Notably, expected gains can be negative when products from the chapter are located far from EV specific products. For jewels \& coins (HS chapter 71) and aircraft \& spacecraft (HS chapter 88), we observe negative expected relative gains of $-0.98$ and $-0.96$, respectively.

These results take path dependence into account and are therefore country-specific. For example, China's new comparative advantage in vehicles yields 1.6 times more expected new EV strengths than the average EV-related sector because China is already diversified. China yields the highest gains in EV strengths from aluminum products, at 4.5. A new comparative advantage in rubber products yields 1.7 new EV strengths.

We do not find a significant association between the EVCP (at the country-product level) and the EV potential in 2012 and the switch to sRCA > 0 from 2012 to 2022 of EV specific products, see Table~\ref{tab:regr_ev}.
The results for the EVCP for product spaces derived from firm- and industry-level data are shown in Figure~\ref{fig:evcpscatter}.

Figures\ref{fig:mealyadapt_ml} and ~\ref{fig:mealyadapt} in the supplementary information show these results in more detail.

The results are presented only for countries that already have a significant number of companies (more than 150), in order to ensure the robustness of the measurement of these values.
Furthermore, while the EVCP indicator weights products proportionally to their complexity indicator, the proximity centrality gives equal weight to each product.

In the supplementary information, Figure~\ref{fig:cninfluences} shows in more detail the chapters per country that contribute most to closeness centrality overall. We find particularly machinery and electronical equipment to be among the highest contributing chapters across all countries, followed by organic chemicals and iron \& steel products as well as optical, medical \& precision instruments. However, the extent to which they contribute is heterogeneous.
Electrical equipment contributes most in countries including Czech Republic, and Mexico but also Thailand, Malaysia and the Philippines.
Machinery contributes most in the Switzerland, Sweden, Italy, Germany, and several other European countries.
We additionally show the groups of products that contribute most to closeness centrality of the electronical equipment category (Figure~\ref{fig:cninfluences85}). The categories that are closest to EV components across all countries comprise electrical ignition or starting equipment, electric motors and generators, electrical lighting or signalling equipment, as well as devices used in circuits exceeding a voltage of 1000 volts.
A similar observation can be made when considering the results for machinery, which are dominated by heating/cooling machinery, pumps, centrifuges, and other parts (Figure~\ref{fig:cninfluences84}).

Further results for unspecific and ICE-specific products can be found in the supplementary information  (Figure~\ref{fig:closeness_unspec_detail} and ~\ref{fig:closeness_ice_detail}).

The results for the EVCP are shown in Figure~\ref{fig:evcpscatter}A for firm-level data ($y$ axis) against the EVCP for industry-level data ($x$ axis).
While Germany ranks highest in EVCP using industry-level trade data, results change on a firm-level.
Switzerland, Korea, Canada, the Netherlands, and China show higher firm-level EVCP values.
Conversely, countries like Poland, Vietnam, and Portugal rank lower on a firm- than industry-level.

Figure~\ref{fig:evcpscatter}B shows the results for EV-specific closeness centrality on a firm-level ($y$ axis) against results obtained on an industry-level ($x$ axis).
For the latter, the highest diversification potential is identified for Switzerland, Hungary, Sweden, Slovakia, and Portugal.
Conversely, countries such as Russia, South Africa, Brazil, Indonesia and India exhibit comparatively lower diversification potential.
Using only the firm-level data, the highest values are observed for Korea, China, USA, Canada, Japan, and UK, followed by France, Germany, Italy and Austria.

\begin{figure}
    \centering
    \includegraphics[width=0.8\linewidth]{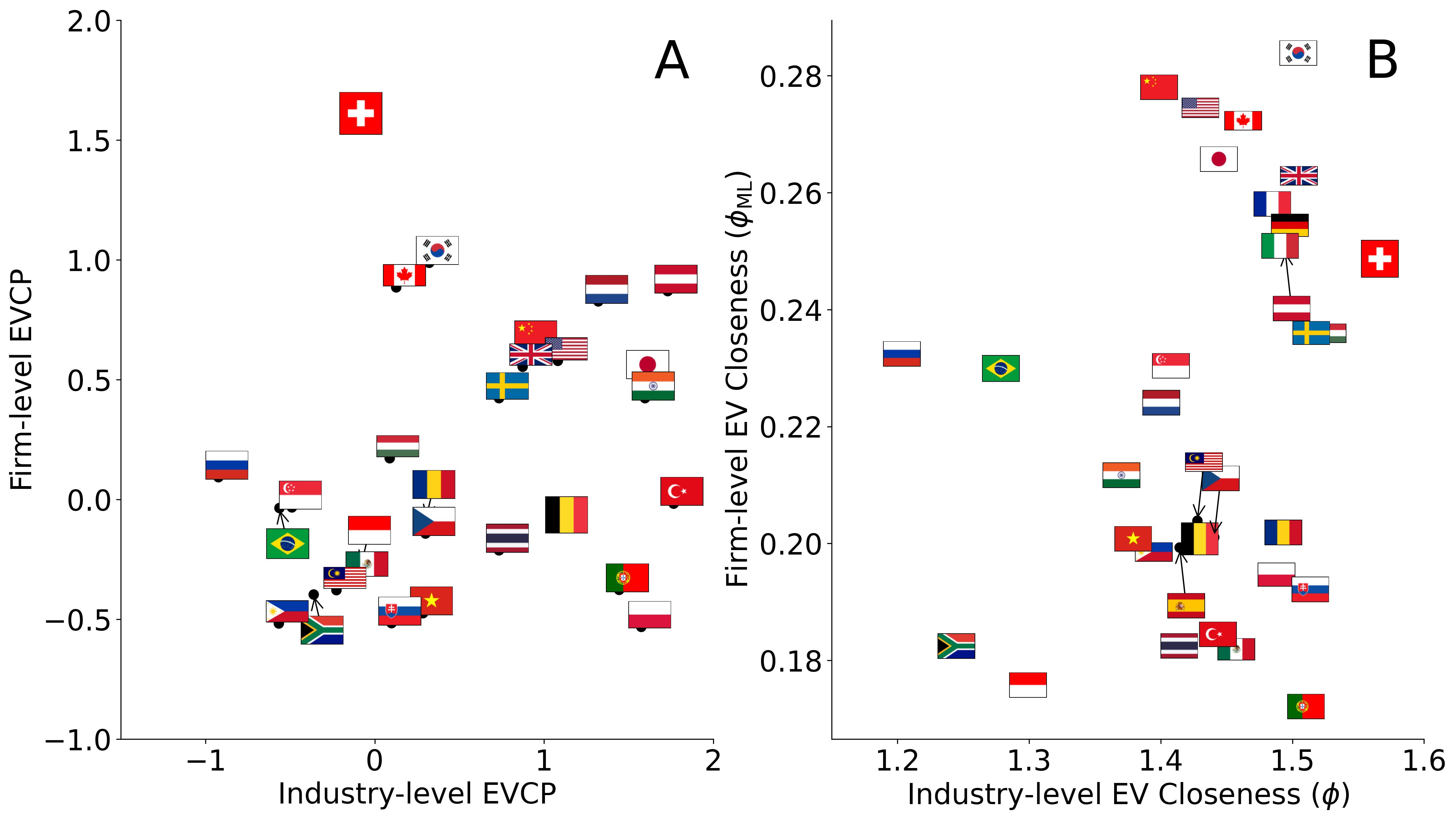}
    \caption{ A) EVCP based on firm-level data and EVCP based on industry-level data. B) Closeness centrality computed from the firm-level product space and from the product spaced constructed from the industry-level product space.}
    \label{fig:evcpscatter}
\end{figure}

\subsection{Opportunities and derisking}

\begin{figure}
    \centering
    \includegraphics[width=.99\linewidth]{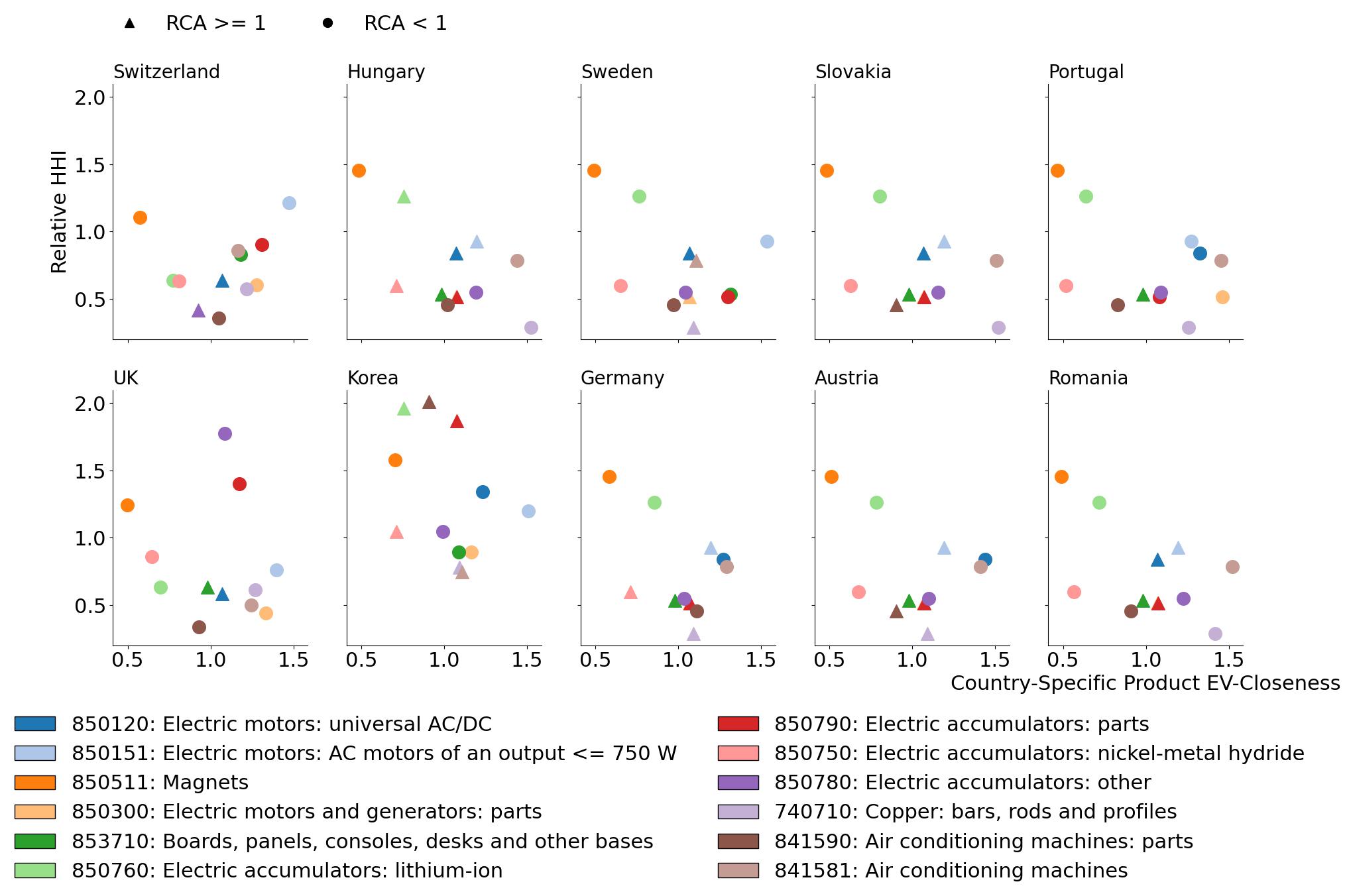}
    \caption{Herfindahl–Hirschman Index (HHI) and closeness of country specific products (RCA >= 1) with EV specific HS codes from the product space constructed from the industry-level product space. Triangles indicate an RCA >= 1 and circles indicate an RCA < 1. }
    \label{fig:hhi}
\end{figure}

The potential for derisking is quantified for individual product categories by means of their relative HHI.
A value of above (below) one for a product in a country is indicative of the fact that the country's imports in this product are more (less) concentrated compared to the average import concentration.
The outcomes of this analysis are presented in Figure ~\ref{fig:hhi}, wherein the relative HHI is plotted against the country-specific closeness centrality of individual EV product categories.
The symbols indicate whether the country already has (triangle) or does not have (circle) comparative advantage in this product.
The relative HHI values of the entire EU Single Market were used for EU countries.

We find that, for the majority of European countries, products with higher relative HHI scores (e.g., permanent magnets, electrical accumulators) exhibit a comparatively low closeness centrality in products that do not represent their comparative advantages (negative correlation in the plots).
These countries are confronted with a trade-off between derisking and leveraging growth opportunities that are more aligned with the existing export portfolio.

We also analyze this for unspecified and ICE components in the supplementary information and find that the results are slightly different (Figures~\ref{fig:hhi_ice} and \ref{fig:hhi_unspec}). While there is also a clear trend that products with a high closeness centrality are also comparative advantages of a country, our analysis shows that there are a number of products with low RCA, high HHI and high closeness centrality in both categories.
Hence, these categories present candidates for diversification that could simultaneously reduce supply chain dependencies.

\section{Discussion}

This study examines the structural transformation of the automotive industry using a component-based approach that combines firm-level and trade data.
By comparing product spaces from these data sources, we can systematically identify opportunities for diversification and de-risking across regions.
Our analysis shows that these opportunities are highly heterogeneous across regions and strongly depend on existing comparative advantages in other industries.

We validate the predictive accuracy of the principle of relatedness by assessing how well sector-specific comparative advantages in close product-space-proximity explain the emergence of new comparative advantages in general and in the EV value chain in particular. Our results show that while proximity is indeed relevant, it is the closeness centrality of a product — its structural position in the global product space — that is more strongly linked to future diversification success. In particular, products that are central (i.e., highly connected to many others) yield greater spillover potential when added to a country’s export basket.

We use sector-specific, sRCA measures to assess a country's export competitiveness in particular sectors, rather than across the entire product range. This approach controls for possible distortions of the comparisons due to sector size or export volume. When analyzing the entire product space, we find that complexity potential and overall potential have larger coefficients than closeness in explaining the emergence of new competitive strengths (with a similar level of statistical significance to that found for closeness centrality). This suggests that they capture broader capability- and opportunity-based dimensions of diversification. 
However, when we restrict sRCA to the sectoral level and focus only on exports within the EV product group, closeness becomes the most significant and consistent predictor, while the potential-based metrics lose both magnitude and significance. This highlights the particular value of proximity-based measures in modeling within-sector upgrading processes, despite the fact that broader capability development can also be captured in the full product space.

For EV-specific products, we find that a one-standard-deviation increase in closeness is associated with a 52\% increase in the likelihood of transitioning from non-competitive to competitive export status (sRCA > 0). In contrast, the complexity potential, which includes product complexity and relatedness to existing capabilities, shows no significant association with export emergence. The same applies to relatedness potential, which omits product complexity.
Country- and sector-specific effects reveal asymmetries. For the EU, new strengths in vehicles (87), aluminum products (76), and rubber (40) result in EV-specific strengths that are 5.0, 4.6, and 3.7 times greater than those in an average EV-related sector. Strengths in other industries that are farther apart in the product space would lead to reductions in EV-specific strengths. This suggests that not all diversification paths are equally beneficial for expanding EV-specific capabilities -- some may divert focus or resources away from strategically aligned trajectories.

We find that the diversification potential for EV components needs to be assessed from different perspectives.
If this potential is assessed on the basis of the comparative advantage of firms within the country that are part of the automotive industry, the top-ranked countries include highly diversified regions with a long tradition in the industry, such as the US, China, Korea, Japan, the UK and many EU countries.
This is the case whether we consider rankings that give more weight to highly complex products (EVCP) or not (closeness centrality).

However, the results change when we assess cross-industry transformation paths and take into account the potential for diversification from other industries. They reflect a country's structural proximity to electric vehicle (EV)-related products based on its overall export composition rather than its existing automotive industry. Countries such as Switzerland, Italy, and Sweden benefit from their machinery sectors, while countries such as Vietnam, the Philippines, Indonesia, and Malaysia gain significant diversification potential from their comparative advantage in electronical equipment.
The structural transformation of the automotive industry could therefore favour new regions, namely those that have emerged in recent years as manufacturing hubs for other industries and are therefore pre-adapted to the production of EV components.

The discrepancy between firm- and industry-level results highlights two complementary dimensions of diversification potential: existing capacity for within-sector upgrading and broader structural readiness for cross-sectoral transition. While firm-level data captures the potential of existing firms active in the automotive supply chain, industry-level indicators reveal latent capabilities in related industries.

The case for analyzing different levels of observation is based on the role of market mechanisms. The capabilities of firms in the auto industry are the starting point for change within existing structures. At a higher level of aggregation, the market level, additional mechanisms become visible that directly shape individual firm performance. This reflects the self-organizing processes highlighted in evolutionary economics \cite{dosi2017footprint, friesenbichler2020high}. In other words, the system-wide evolution of industries cannot be deduced from their individual components (here: firms) alone. Instead, aggregate outcomes emerge from off-equilibrium interactions of economic agents, intra- and extra-industry selection, and heterogeneous learning processes \cite{dosi2005statistical}.

We also show that many EU countries not only lack comparative advantages and diversification potential for EV components compared to other countries, but that this is combined with high import concentration and thus potential supply chain vulnerabilities.
This could have far-reaching implications for EU industrial policy.
Components such as EV batteries have become strategic goods whose production is heavily subsidized in China and the US.
Given the apparent lack of pre-adapted industries in the EU, it is questionable whether EU companies can catch up with the new market leaders in developing an autonomous EV battery value chain.

However, our analysis also reveals that the EU is much better positioned to further diversify into and strengthen its position in other car components.

\subsection{Limitations and considerations}

Our study has several limitations related to data availability and methodological approach.
First, the firm-level data from the Marklines supplier database presents challenges. As a proprietary dataset based on self-reported manufacturer data, its coverage is limited to firms willing to disclose their supply chain relationships. The classification of components into EV-specific, ICE-specific, and non-specific categories was done manually, as was the mapping of Marklines components to HS codes.
Even though these mapping were performed by a team of experienced industry analysts, with each relationship being checked by at least two researchers, inconsistencies and omissions cannot be fully ruled out.
Furthermore, the dataset focuses exclusively on final automotive components, excluding upstream processes such as molding, welding, or raw material extraction, which are crucial to the production ecosystem.

Second, the BACI trade dataset, based on UN Comtrade data, has additional limitations. Trade data do not always match actual production locations. For example, US-owned factories operating in China or Chinese-owned manufacturing plants in Mexico lead to discrepancies in identifying where technological capabilities are located. Some countries act as intermediaries in global value chains, assembling components manufactured elsewhere without significant local production. This distorts export-based assessments of industrial capabilities. Furthermore, export data only serve as a proxy for underlying technological and production capabilities. It does not capture all the factors that contribute to industrial strength, such as domestic market dynamics, patented innovations, and foreign direct investment. For example, the construction of Tesla's Gigafactory in Germany raises the question of whether this production capacity should be attributed to the development of local capabilities or to the technological foundation of a foreign company. Such cases underscore a well-known shortcoming of export-based complexity measures\cite{KoopmanWangWei2012}. Countries that primarily serve as manufacturing hubs, such as Mexico and several Eastern European nations, frequently appear to have disproportionately high complexity scores. This occurs because export-based measures can overstate or understate national capabilities, reflecting the sophistication of exported products, not the depth of domestic technological capability or local value added.

The analysis is further constrained by the resolution of trade data, which is limited to the six-digit level of the HS classification. This granularity may be too coarse to accurately distinguish between EV-specific and ICE-specific components. To address this, HS codes were classified as EV-specific if more than 50\% of the corresponding Marklines products were identified as EV-related. However, this threshold-based classification introduces potential inaccuracies and may not capture finer distinctions between multi-use components. Discrepancies in historical trade data due to shifts from HS07 to HS12 classifications further complicate longitudinal analyses. Moreover, the product space derived from the BACI data was constructed only for 2022, and the Marklines dataset lacks temporal information, preventing an assessment of structural transformations over time.

Finally, while the economic complexity framework systematically quantifies industrial transformation, it does not fully account for policy interventions in shaping industrial development. The principle of relatedness assumes diversification occurs incrementally based on pre-existing capabilities. However, targeted policies, subsidies, and regulatory frameworks can significantly accelerate or impede this process.

For example, China’s rapid expansion of electric vehicle manufacturing is driven not only by relatedness but also by government incentives, local content requirements, and strategic industrial planning. Similarly, the EU’s push into battery manufacturing is influenced by regulatory frameworks rather than purely market-driven capability accumulation. The interaction between policy and capability development is complex. While relatedness metrics provide valuable insights, they do not capture the full range of factors influencing industrial transformation.

Compared to previous complexity-based analyses of the automotive industry, our findings build upon existing work by offering a product-based, cross-industry perspective. For example, Yamada \textit{et al.}~\cite{yamada2024structure} constructed a firm-level product space for Japanese carmakers and demonstrated that diversification adheres to the same network principles observed in international trade data. However, their analysis remained confined to the domestic automotive sector. Bam \textit{et al.}\cite{bam2021io} examined sectoral upgrading in the South African automotive value chain by combining trade data with input-output linkages, capturing diversification at the level of industry aggregates rather than individual products. Cresti \textit{et al.}\cite{cresti2025vulnerabilitiescapabilitieseuautomotive} analyzed the diversification potential and vulnerabilities of the EU automotive supply chain using sector-level trade and input-output data. However, their study does not model cross-industry linkages or predict diversification outcomes because intersectoral dependencies are assessed at the sector level rather than the product level. In contrast, our framework constructs a high-resolution product-based network linking firm-level and trade data to model diversification pathways across industries and evaluate their predictive power. This approach captures how the structural position of products within the global product space shapes a country’s potential to expand into EV-related technologies. It offers a more detailed and predictive view of industrial transformation.

\section{Conclusion}
This paper uses a product-space framework that compares firm-level and trade data to explain the transformation of the automotive industry. This approach enables the study of diversification dynamics within and across industries. We argue that opportunities to diversify into EV components are largely path dependent and concentrated in countries with established strengths in machinery, vehicles, and electrical equipment. 

The analysis is also limited by the scope and granularity of the data available. For example, trade data may not accurately reflect production capacity, and the classification of EV and ICE components relies in part on judgment. Furthermore, the framework captures capability-based transformation, yet it does not consider the political and institutional factors that influence industrial change. Despite these limitations, the results show how combining trade with firm-level information reveals structural constraints and opportunities in technological transitions.

Comparing complexity-based indicators with network-based measures shows that both capture important aspects of capability formation. Yet, the latter better reflects accessible diversification pathways in the context of the EV transition.

This paper departs from a strand of evolutionary economics that describes empirical patterns and statistical regularities of growth processes \cite{dosi2017footprint, hidalgo2009building}, which we break down to a sectoral analysis. Our method is based on export data and draws on the relevance economic structures and path-dependent capabilities that channel development processes. The indicators used in this study are indicative of development paths, but they may materialize differently depending on factors such as industrial policy or unanticipated technological developments. The results should be interpreted against this backdrop. The findings suggest that, for a successful transition, policymakers must build on existing industrial bases while closing gaps in key EV inputs, such as batteries and magnets.

\section*{Conflicts of Interest} 
The authors declare no conflicts of interest. 
\section*{Author Contributions}
PK conceptualized and supervised the project. PK and KL devised the analytic methods. KL carried out the analysis and produced the plots and graphics. KL and PK wrote the first draft of the manuscript. KF contributed to the academic positioning. LB constructed the dataset for the firm-level analysis. 
KF, KL and PK conducted reviewing and editing of the manuscript. All authors read and approved the final manuscript.
\section*{Acknowledgments}
On behalf of the Supply Chain Intelligence Institute Austria (ASCII), we acknowledge financial support from the Austrian Federal Ministry for Economy, Energy and Tourism (BMWET) and the Federal State of Upper Austria.
\printbibliography
\newpage
\section*{Appendix}
\renewcommand\theequation{\Alph{section}\arabic{equation}} 
\counterwithin*{equation}{section} 
\setcounter{figure}{0}
\setcounter{table}{0}
\renewcommand{\figurename}{Figure}
\renewcommand{\thefigure}{S\arabic{figure}}
\counterwithin*{figure}{section} 

\setcounter{table}{0}
\renewcommand{\tablename}{Table}
\renewcommand{\thetable}{S\arabic{table}}

\begin{appendices}


\begin{figure}[h!]
    \centering
    \includegraphics[width=.8\linewidth]{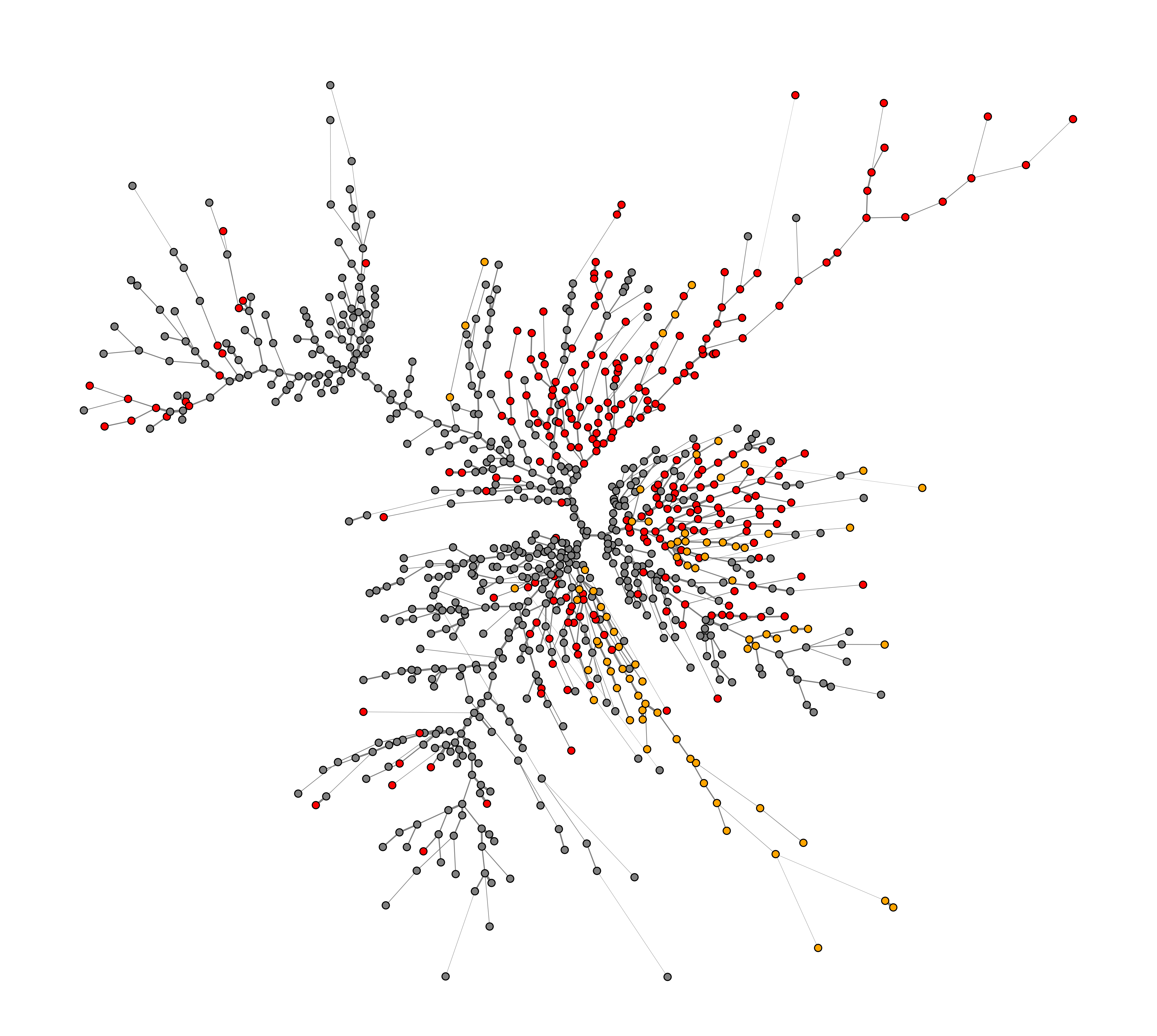}
    \caption{Firm-level product space. Grey nodes are unspecific automotive products, red nodes are ICE-specific products, and orange nodes are EV-specific products.}
    \label{fig:ml_productspace}
\end{figure}
 

\begin{figure}[h!]
    \centering
    \includegraphics[width=.8\linewidth]{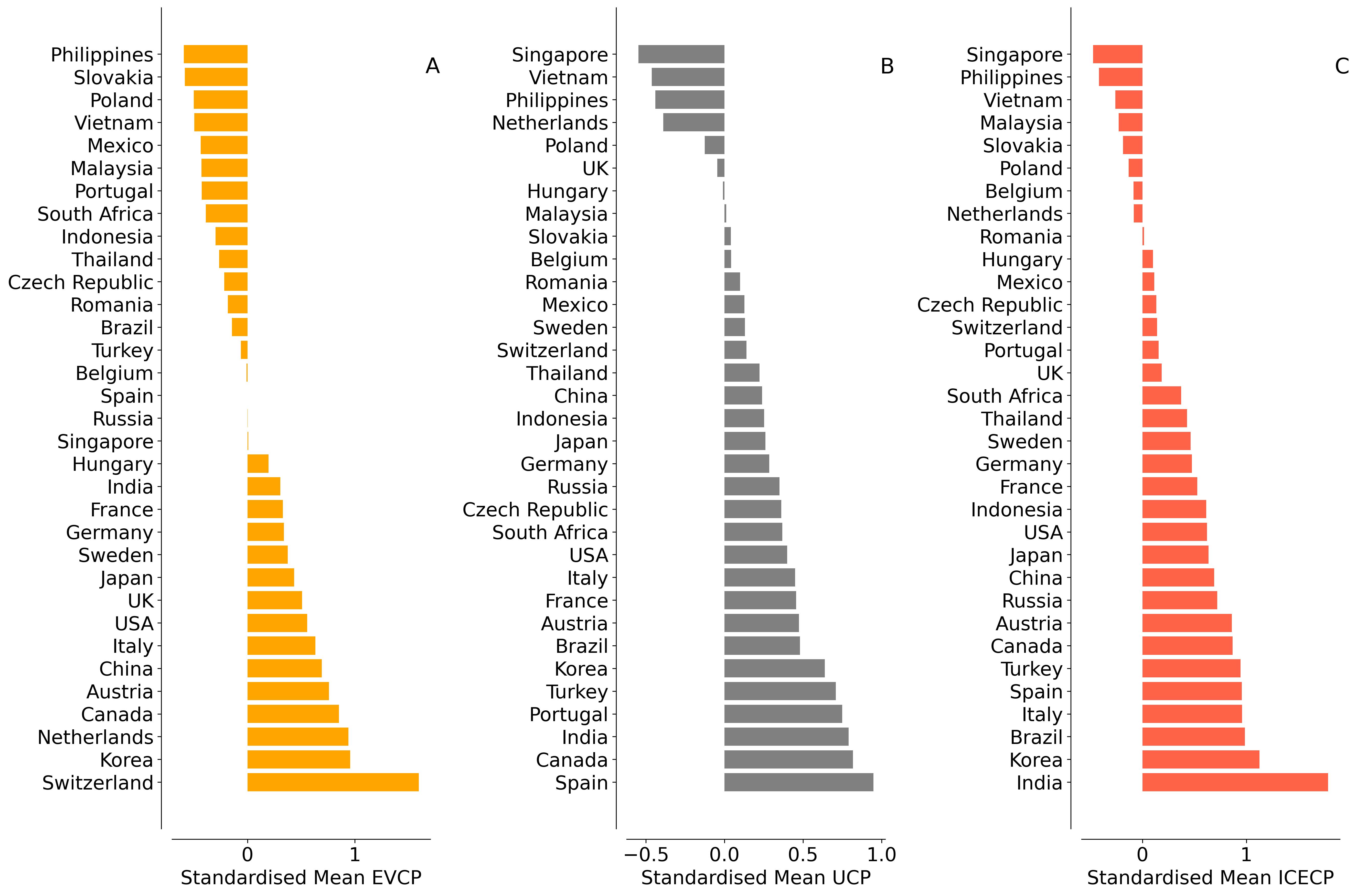}
    \caption{EV-Complexity Potential (EVCP) (A), Unspecific-Complexity Potential (UCP) (B), and ICE-Complexity Potential (C) based on Marklines dataset for countries with more than 150 firms in Marklines.}
    \label{fig:mealyadapt_ml}
\end{figure}

\begin{figure}[h!]
    \centering
    \includegraphics[width=.8\linewidth]{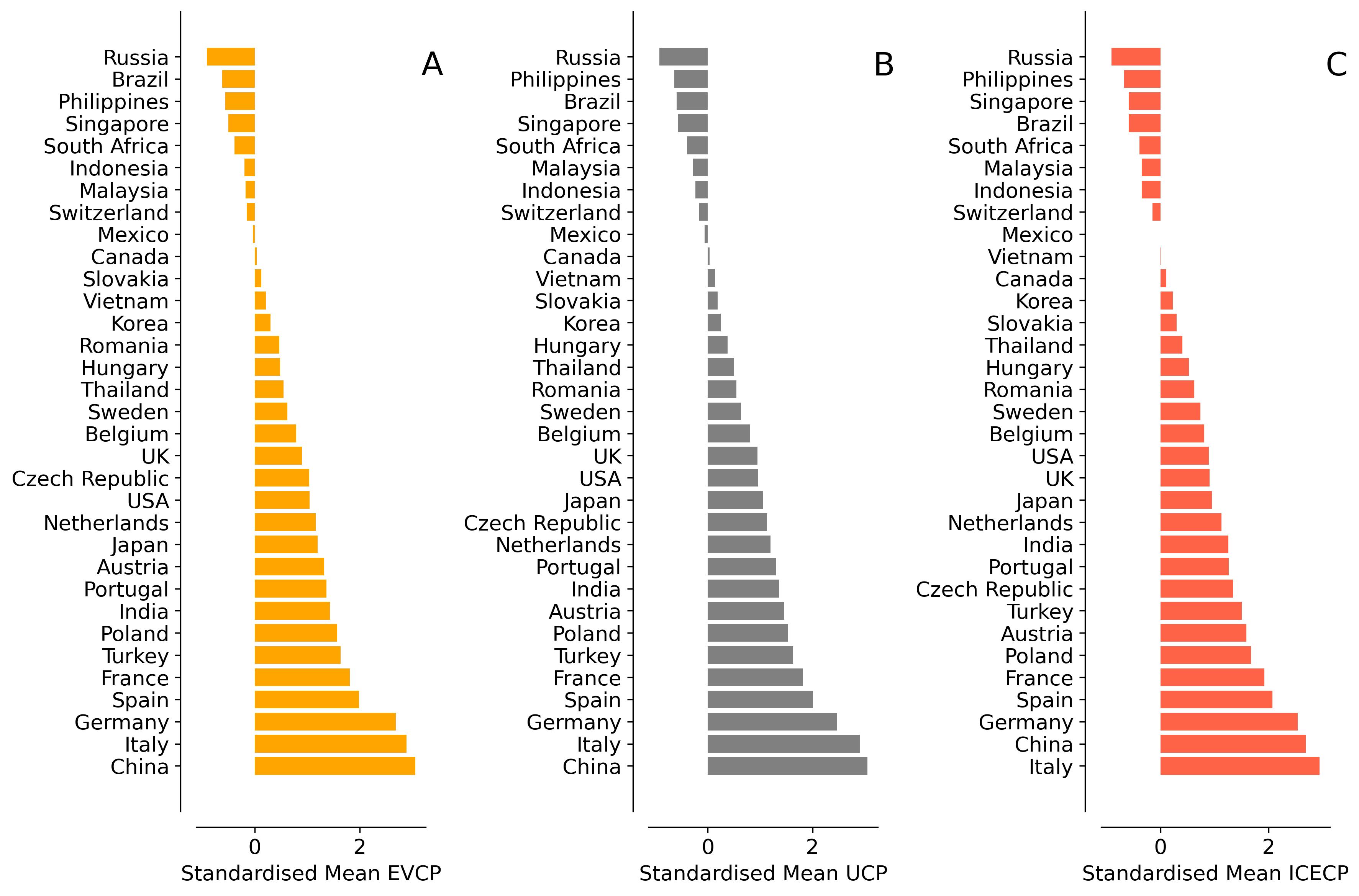}
    \caption{EV-Complexity Potential (EVCP) (A), Unspecific-Complexity Potential (UCP) (B), and ICE-Complexity Potential (C) based on BACI dataset (2022) for countries with more than 150 firms in Marklines.}
    \label{fig:mealyadapt}
\end{figure}
\begin{figure}
    \centering
    \includegraphics[width=0.85\linewidth]{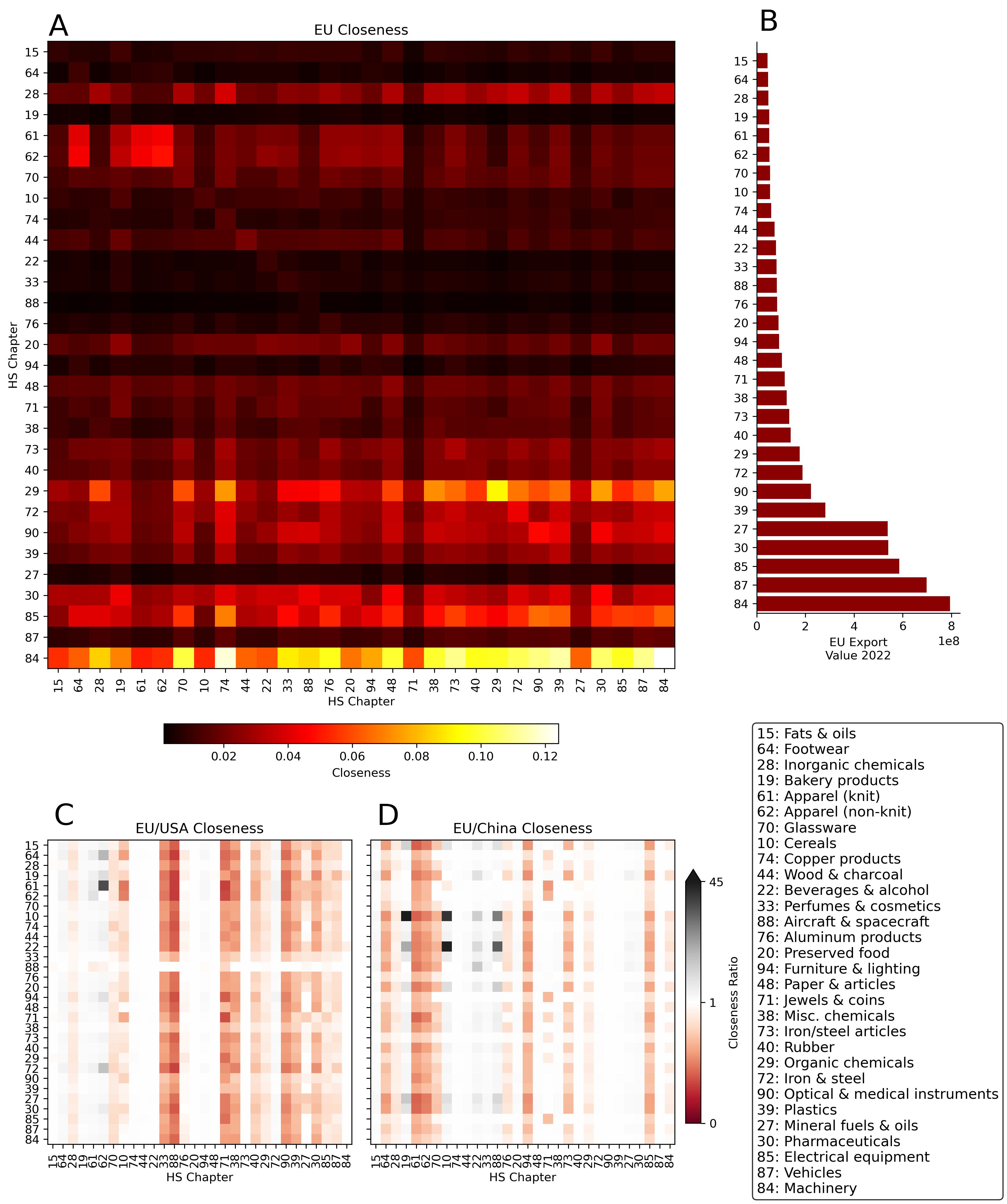}
    \caption{Closeness-Metric of EU Export Specialization at the Industry Level. A) Normalized closeness between the 30 HS chapters with the highest EU export value in 2022, measured as the average pairwise closeness between products in each chapter pair (i.e., sum over $C_{ij}$ normalized by $N_j$.) B) Total EU export value in 2022 for the top 30 HS chapters. C) Closeness ratio of the EU relative to the USA (EU / USA) and D) closeness ratio of the EU relative to China.}
    \label{fig:closeness_baci_descriptive}
\end{figure}

\begin{table}[ht]
\centering
\begin{tabular}{lccc}
\toprule
Chapter & $C_p$ & $CP_p$ & $P_p$ \\
\midrule
10 & 0.295 & -0.283 & -0.401 \\
15 & 0.543*** & 0.333 & 0.545 \\
19 & 0.364 & -0.548*** & -0.490* \\
20 & 0.482** & -0.268 & -0.163 \\
22 & 0.035 & -0.081 & -0.065 \\
27 & 0.489*** & -0.108 & -0.152 \\
28 & 0.168 & -0.254 & -0.484*** \\
29 & 0.720*** & 0.003 & 0.138 \\
30 & 0.465*** & -0.289* & -0.265 \\
33 & 0.091 & 0.032 & -0.045 \\
38 & 0.034 & -0.153 & -0.146 \\
39 & -0.196 & 0.009 & 0.012 \\
40 & 0.427*** & -0.318** & -0.269 \\
44 & 0.032 & -0.053 & -0.046 \\
48 & 0.165 & -0.257 & -0.195 \\
61 & -0.047 & -0.126 & -0.075 \\
62 & -0.179 & -0.402** & -0.483*** \\
64 & 0.182 & -0.004 & 0.021 \\
70 & 0.402** & -0.316** & -0.201 \\
71 & 0.145 & -0.236 & -0.219 \\
72 & 0.263 & -0.045 & -0.164 \\
73 & 0.076 & -0.357** & -0.343** \\
74 & 0.197 & -0.004 & -0.035 \\
76 & 0.065 & -0.310** & -0.299** \\
84 & 0.206 & -0.272 & -0.262 \\
85 & 0.643*** & 0.519** & 1.046*** \\
87 & 0.180 & -0.081 & -0.109 \\
88 & 0.124 & 0.525*** & 0.366** \\
90 & 0.600*** & 0.384** & 0.405** \\
\bottomrule
\end{tabular}
\caption{Regression results per HS chapter for 12 randomly selected products. Significance levels: * $p<0.05$, ** $p<0.01$, *** $p<0.001$}
\label{tab:regr_random}
\end{table}

\begin{table}[ht]
\centering
\begin{tabular}{lccc}
\toprule
Chapter & $C_p$ & $CP_p$ & $P_p$ \\
\midrule
10 & -0.120 & -0.280 & -0.326 \\
15 & 0.672*** & 0.648 & 0.700* \\
19 & 0.488* & -0.368 & -0.348 \\
20 & 0.510*** & 0.123 & 0.051 \\
22 & 0.416* & -0.417 & -0.376 \\
27 & -0.049 & -0.168 & -0.109 \\
28 & 0.552*** & 0.994 & 1.320 \\
29 & 0.093 & -0.546* & -0.515* \\
30 & 0.224 & -0.609** & -0.615** \\
33 & 0.391** & 0.049 & 0.015 \\
38 & 0.081 & 0.269 & 0.256 \\
39 & 0.095 & -0.709*** & -0.706*** \\
40 & 0.202 & -0.210 & -0.207 \\
44 & 0.060 & -0.236 & -0.253 \\
48 & 0.614*** & -0.298 & -0.318* \\
61 & -0.151 & -0.196 & -0.227* \\
62 & -0.319** & -0.097 & -0.113 \\
64 & 0.053 & -0.016 & -0.013 \\
70 & 0.353* & -0.247 & -0.227 \\
71 & 0.464*** & 0.379 & 0.441 \\
72 & 0.365* & -0.436* & -0.449** \\
73 & 0.036 & -0.340* & -0.341* \\
74 & 0.155 & 0.110 & 0.133 \\
76 & 0.276 & -0.480** & -0.463** \\
84 & 0.200 & -0.256 & -0.248 \\
85 & 0.132 & -0.205 & -0.187 \\
87 & 0.120 & -0.212 & -0.215 \\
\bottomrule
\end{tabular}
\caption{Regression results per HS chapter for 12 randomly selected products from those in the top quartile of Product Complexity Index (PCI). Significance levels: * $p<0.05$, ** $p<0.01$, *** $p<0.001$}
\label{tab:regr_toppci}
\end{table}

\begin{table}[ht]
\centering
\begin{tabular}{lccc}
\toprule
Chapter & $C_p$ & $CP_p$ & $P_p$ \\
\midrule
10 & 0.243 & 0.061 & 0.028 \\
15 & 0.703*** & 0.392* & 0.360 \\
19 & 0.265 & -0.174 & -0.121 \\
20 & 0.408* & 0.279 & 0.347* \\
22 & 0.120 & -0.129 & -0.116 \\
27 & 0.366 & 0.149 & 0.137 \\
28 & 0.460** & 0.228 & 0.215 \\
29 & 0.316 & -0.123 & -0.105 \\
30 & 0.151 & -0.081 & -0.063 \\
33 & 0.434** & 0.193 & 0.174 \\
38 & 0.080 & -0.065 & -0.055 \\
39 & 0.202 & 0.131 & 0.117 \\
40 & 0.140 & -0.054 & -0.049 \\
44 & 0.269 & 0.035 & 0.024 \\
48 & 0.347 & 0.198 & 0.188 \\
61 & -0.043 & -0.129 & -0.116 \\
62 & -0.011 & -0.204 & -0.193 \\
64 & 0.036 & -0.112 & -0.094 \\
70 & 0.370* & -0.014 & -0.034 \\
71 & 0.422** & 0.005 & -0.001 \\
72 & 0.273 & -0.239 & -0.259 \\
73 & 0.148 & -0.010 & -0.007 \\
74 & 0.236 & 0.081 & 0.071 \\
76 & 0.143 & -0.181 & -0.178 \\
84 & 0.366* & -0.137 & -0.146 \\
85 & 0.388** & 0.078 & 0.057 \\
87 & 0.238 & 0.056 & 0.054 \\
\bottomrule
\end{tabular}
\caption{Regression results per HS chapter for 12 randomly selected products from those in the bottom quartile of Product Complexity Index (PCI). Significance levels: * $p<0.05$, ** $p<0.01$, *** $p<0.001$}
\label{tab:regr_botpci}
\end{table}

\begin{table}[ht]
\centering
\begin{tabular}{lccc}
\toprule
Chapter & $C_p$ & $CP_p$ & $P_p$ \\
\midrule
EV & 0.419*** & 0.145 & 0.143 \\
\bottomrule
\end{tabular}
\caption{Regression results for EV-specific products. Significance levels: * $p<0.05$, ** $p<0.01$, *** $p<0.001$}
\label{tab:regr_ev}
\end{table}

\begin{table}
\centering
\begin{tabular}{lccc}
\toprule
Chapter & $C_p$ & $CP_p$ & $P_p$ \\
\midrule
10 & -0.061 & 0.811 & 2.588** \\
15 & 0.392** & 0.874** & 0.872 \\
19 & 0.756*** & 1.160** & 3.323*** \\
20 & 0.273 & 0.549 & 1.252* \\
22 & 0.502*** & 0.237 & 1.623** \\
27 & 0.345* & 0.454 & 0.519 \\
28 & 0.412** & 1.232* & 0.709 \\
29 & 0.278 & 0.718 & 0.890 \\
30 & 0.182 & 0.647 & 0.488 \\
33 & 1.052*** & 2.040*** & 3.446*** \\
38 & 0.193 & 0.169 & 0.830 \\
39 & 0.618*** & 0.781 & 1.108 \\
40 & 0.410* & 0.947 & 0.949 \\
44 & 0.825*** & 2.413*** & 3.157*** \\
48 & 0.711*** & 1.411** & 2.155*** \\
61 & 1.144*** & 2.419*** & 3.055*** \\
62 & 1.058*** & 2.412*** & 2.890*** \\
64 & 0.774*** & 1.363** & 1.641** \\
70 & 0.542*** & 1.165* & 1.191* \\
71 & 0.566*** & 1.518** & 1.432** \\
72 & 0.170 & 0.331 & 0.437 \\
73 & 0.687*** & 1.660** & 2.809*** \\
74 & 0.141 & 0.805 & 1.026 \\
76 & 0.717*** & 2.180*** & 2.388*** \\
84 & 0.692*** & 2.014*** & 2.498*** \\
85 & 1.236*** & 2.811*** & 3.046*** \\
87 & 0.579*** & 0.419 & 1.260* \\
88 & 0.710*** & 2.057*** & 2.469*** \\
90 & 0.598*** & 1.756*** & 1.836*** \\
94 & 0.918*** & 2.870*** & 3.263*** \\
\bottomrule
\end{tabular}
\caption{Regression results per HS chapter for 12 randomly selected products using sRCA on all products. Significance levels: * $p<0.05$, ** $p<0.01$, *** $p<0.001$}
\label{tab:regr_random_fullsrca}
\end{table}

\begin{table}
\centering
\begin{tabular}{lccc}
\toprule
Chapter & $C_p$ & $CP_p$ & $P_p$ \\
\midrule
10 & 0.212 & 1.222** & 1.291** \\
15 & 0.585*** & 1.880*** & 1.803*** \\
19 & 1.106*** & 3.027*** & 3.314*** \\
20 & 0.700*** & 2.576*** & 3.107*** \\
22 & 0.018 & 2.339** & 1.860 \\
27 & 0.812*** & 1.340** & 1.169* \\
28 & 0.039 & 0.098 & 0.023 \\
29 & 0.349 & 0.775 & 0.939 \\
30 & 0.248 & 2.435** & 2.532** \\
33 & 1.320*** & 4.673*** & 4.640*** \\
38 & -0.050 & 0.193 & 0.514 \\
39 & 0.493** & 2.603*** & 2.557*** \\
40 & 0.496** & 1.106* & 1.021* \\
44 & 0.547*** & 3.527*** & 3.215*** \\
48 & 0.507*** & 0.141 & 0.203 \\
61 & 0.884*** & 2.679*** & 2.740*** \\
62 & 1.142*** & 3.355*** & 3.237*** \\
64 & 0.445** & 1.327* & 1.428* \\
70 & 0.706*** & 1.709*** & 1.727*** \\
71 & 0.284 & 1.503** & 1.529** \\
72 & -0.006 & 0.796 & 0.852 \\
73 & 0.313 & 1.642* & 1.800* \\
74 & 0.150 & 2.089** & 1.936** \\
76 & 0.236 & 0.091 & 0.064 \\
84 & 0.624*** & 0.536 & 0.482 \\
85 & 0.889*** & 1.977*** & 1.901** \\
87 & 0.239 & 1.078 & 1.009 \\
88 & 0.829 & 2.946 & 2.832 \\
90 & 0.510*** & 1.793** & 1.736** \\
94 & 0.682*** & 2.501*** & 2.805*** \\
\bottomrule
\end{tabular}
\caption{Regression results per HS chapter for 12 randomly selected products using sRCA on all products from those in the top quartile of Product Complexity Index (PCI). Significance levels: * $p<0.05$, ** $p<0.01$, *** $p<0.001$}
\label{tab:regr_toppci_fullsrca}
\end{table}

\begin{table}
\centering
\begin{tabular}{lccc}
\toprule
Chapter & $C_p$ & $CP_p$ & $P_p$ \\
\midrule
10 & 0.614** & 3.071*** & 3.110*** \\
15 & 0.849*** & 3.046*** & 2.762*** \\
19 & 0.660*** & 2.619*** & 2.952*** \\
20 & 0.224 & 1.213 & 1.618 \\
22 & 0.564* & 2.466** & 2.561* \\
27 & -0.065 & 1.664 & 1.376 \\
28 & 0.345 & 0.194 & 0.111 \\
29 & -0.518 & 0.030 & -0.051 \\
30 & 0.165 & 0.265 & 0.371 \\
33 & 0.276 & 0.869 & 1.042 \\
38 & -0.281 & 0.565 & 1.011* \\
39 & 0.390** & 2.030*** & 2.030*** \\
40 & 0.538*** & 1.766*** & 1.835*** \\
44 & 0.578*** & 2.281*** & 2.217*** \\
48 & 0.802*** & 2.792*** & 2.832*** \\
61 & 1.038*** & 1.081*** & 2.726*** \\
62 & 0.830*** & 2.662*** & 2.833*** \\
64 & 0.679*** & 2.141*** & 2.193*** \\
70 & 0.490** & 1.778*** & 2.008*** \\
71 & -0.039 & 1.616*** & 1.545*** \\
72 & 0.231 & -0.255 & -0.287 \\
73 & 0.575*** & 1.833** & 1.871** \\
74 & -0.040 & 1.122 & 1.416* \\
76 & 0.413** & 1.792** & 1.670** \\
84 & 0.474** & 1.600** & 1.611** \\
85 & 0.328* & 1.575** & 1.529** \\
87 & 0.139 & -0.080 & 0.150 \\
88 & 0.341 & 1.330 & 1.184 \\
90 & 0.336 & 1.252* & 1.018 \\
94 & 0.705*** & 2.705*** & 2.810*** \\
\bottomrule
\end{tabular}
\caption{Regression results per HS chapter for 12 randomly selected products using sRCA on all products from those in the bottom quartile of Product Complexity Index (PCI). Significance levels: * $p<0.05$, ** $p<0.01$, *** $p<0.001$}
\label{tab:regr_botpci_fullsrca}
\end{table}

\begin{table}
\centering
\begin{tabular}{lccc}
\toprule
Chapter & $C_p$ & $CP_p$ & $P_p$ \\
\midrule
EV & 0.492** & 0.875** & 0.935** \\
\bottomrule
\end{tabular}
\caption{Regression results for EV-specific products using sRCA on all products. Significance levels: * $p<0.05$, ** $p<0.01$, *** $p<0.001$}
\label{tab:regr_ev_fullsrca}
\end{table}

\begin{figure}[h!]
    \centering
    \includegraphics[width=.8\linewidth]{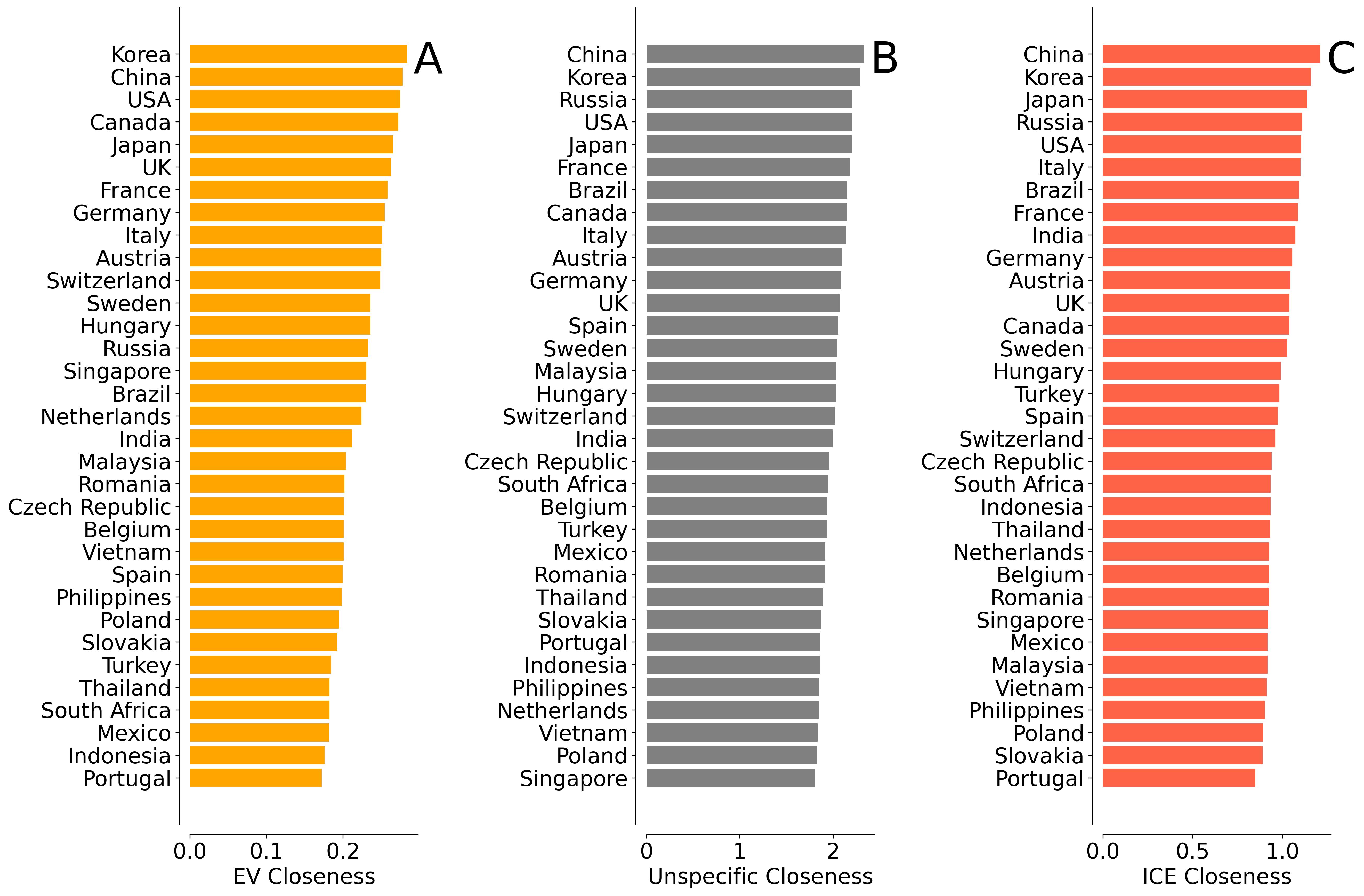}
    \caption{Closeness centrality of country specific products (RCA >= 1) to EV specific products (A), unspecific products (B), and ICE-specific products (C) based on the product space constructed from the Marklines dataset. We show the closeness centrality for countries with more than 150 firms.}
    \label{fig:overlapml}
\end{figure}

\begin{figure}[h!]
    \centering
    \includegraphics[width=.8\linewidth]{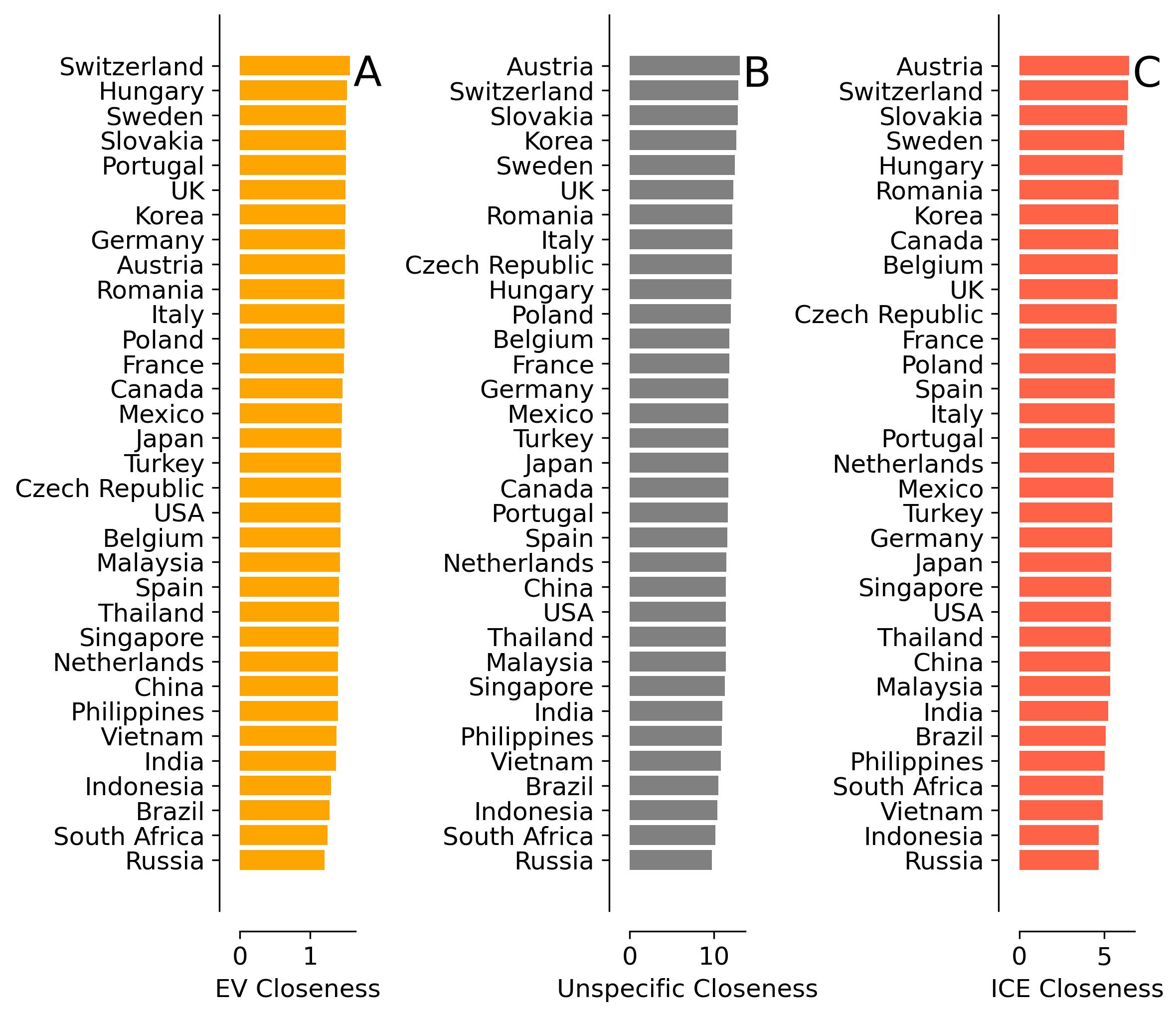}
    \caption{Closeness centrality of country specific products (RCA >= 1) to EV specific products (A), unspecific products (B), and ICE-specific products (C) from the product spaced constructed from the Baci dataset. We show the closeness centrality for countries with more than 150 firms that exist in both datasets.}
    \label{fig:ccbaci}
\end{figure}
\begin{figure}
    \centering
    \includegraphics[width=0.99\linewidth]{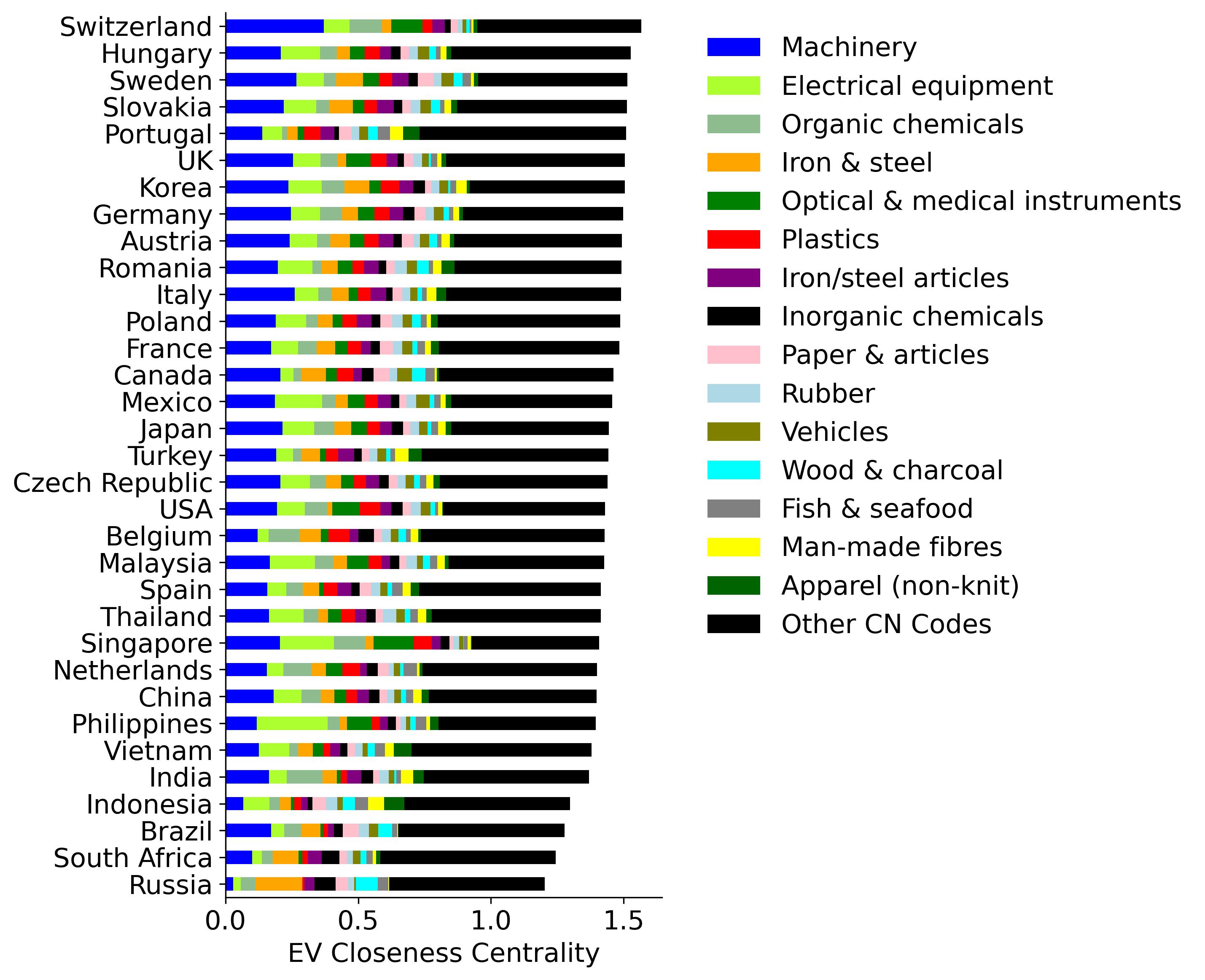}
    \caption{Overview of influences of HS chapters on EV closeness centrality of country specific products (RCA > 1) with EV specific products from the industry-level product space. We show the shares per HS-codes for the 10 most influential HS-codes and the EV closeness centrality for countries with more than 150 firms that exist in both datasets.}
    \label{fig:cninfluences}
\end{figure}
\begin{figure}
    \centering
    \includegraphics[width=.99\linewidth]{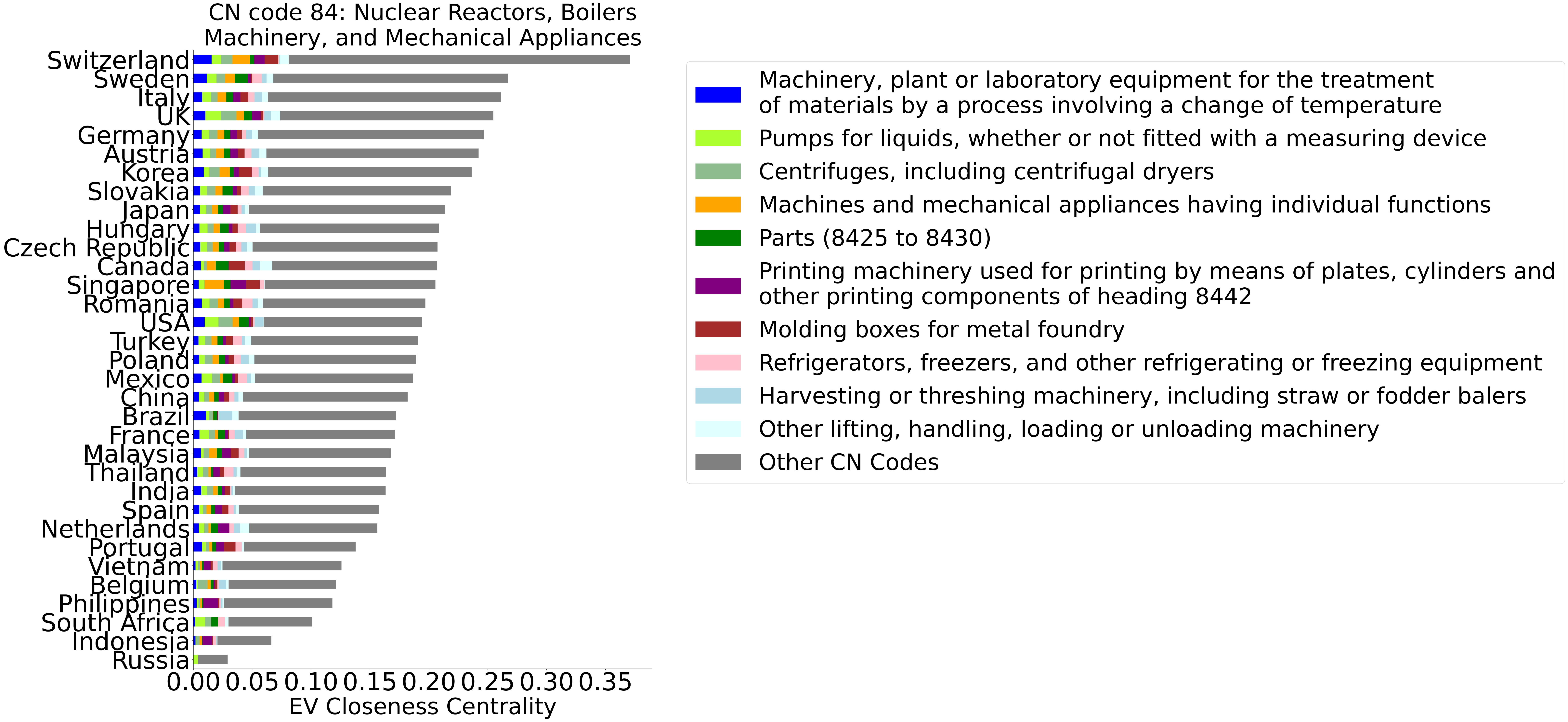}
    \caption{Detailed overview of influences of HS codes 84 on EV closeness centrality of country specific products (RCA > 1) with EV specific products from the industry-level product space. We show the shares per HS-codes for the 10 most influential HS-codes and the EV closeness centrality for countries with more than 150 firms that exist in both datasets.}
    \label{fig:cninfluences84}
\end{figure}

\begin{figure}
    \centering
    \includegraphics[width=.99\linewidth]{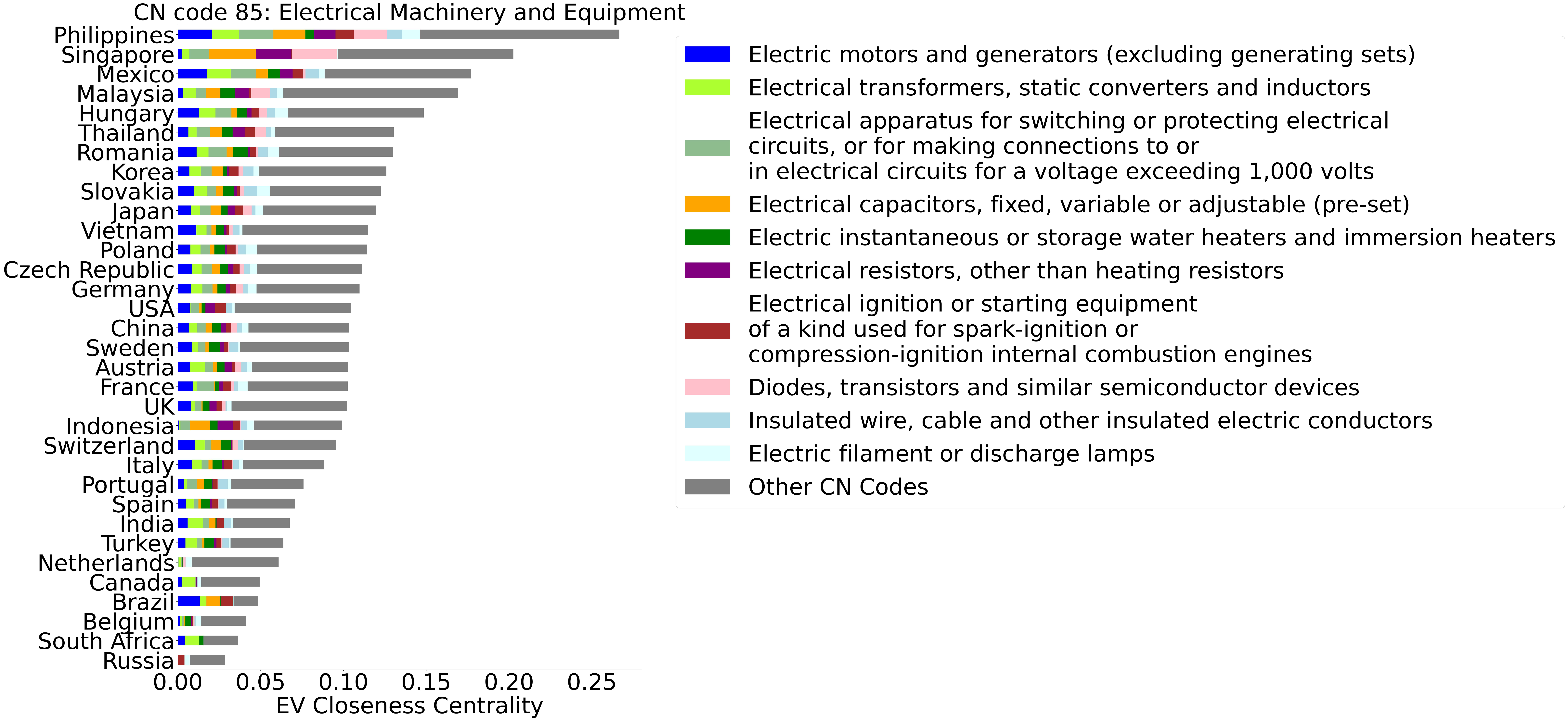}
    \caption{Detailed overview of influences of HS codes 85 on EV closeness centrality of country specific products (RCA > 1) with EV specific products from the industry-level product space. We show the shares per HS-codes for the 10 most influential HS-codes and the EV closeness centrality for countries with more than 150 firms that exist in both datasets.}
    \label{fig:cninfluences85}
\end{figure}
\begin{figure}
    \centering
    \includegraphics[width=0.8\linewidth]{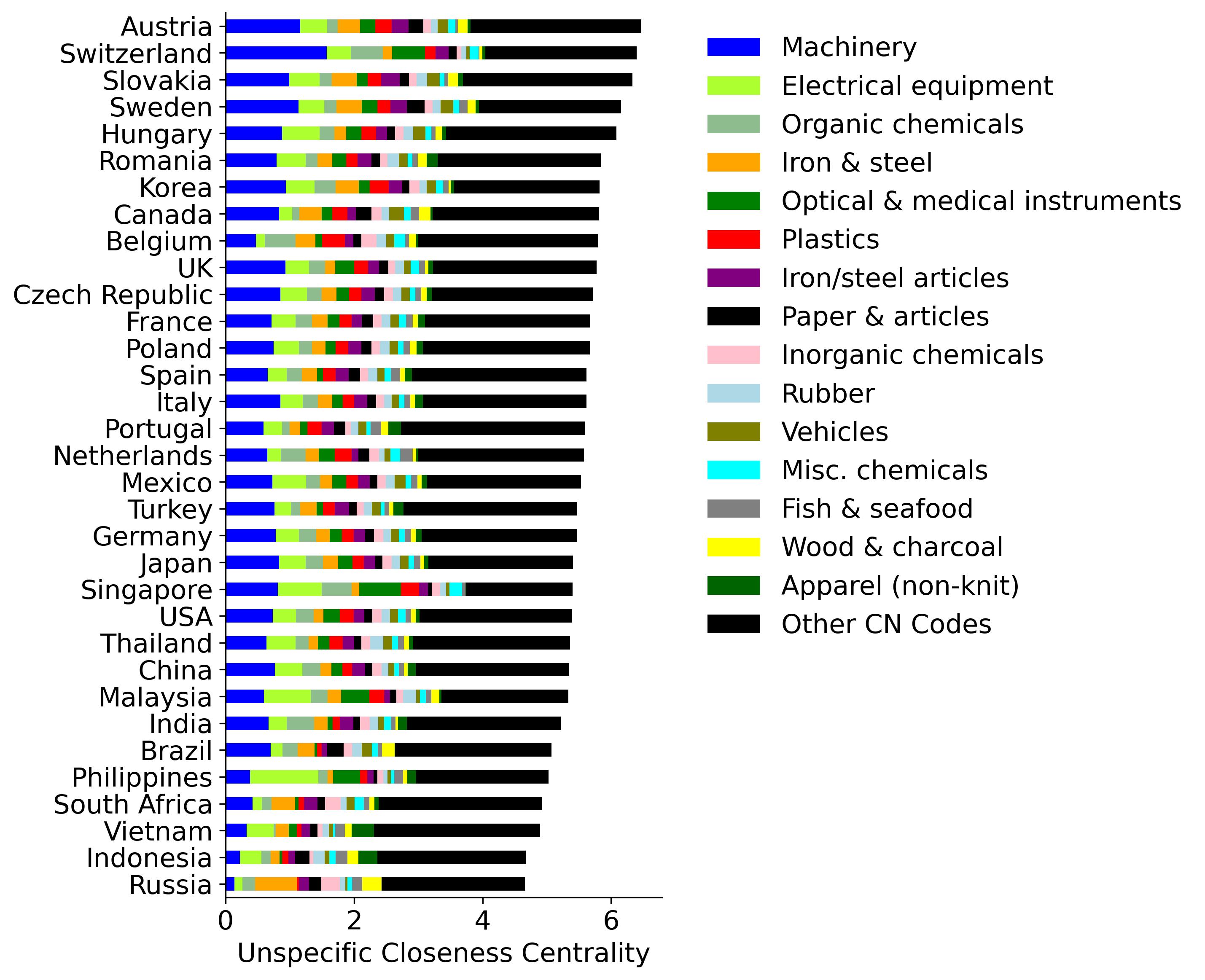}
    \caption{Closeness centrality of country specific products (RCA >= 1) to  Unspecific products from the product spaced constructed from the industry-level product space. We show the shares per HS-codes for the 15 most influential HS-codes for countries with more than 150 firms that exist in both datasets.}
    \label{fig:closeness_unspec_detail}
\end{figure}

\begin{figure}
    \centering
    \includegraphics[width=0.8\linewidth]{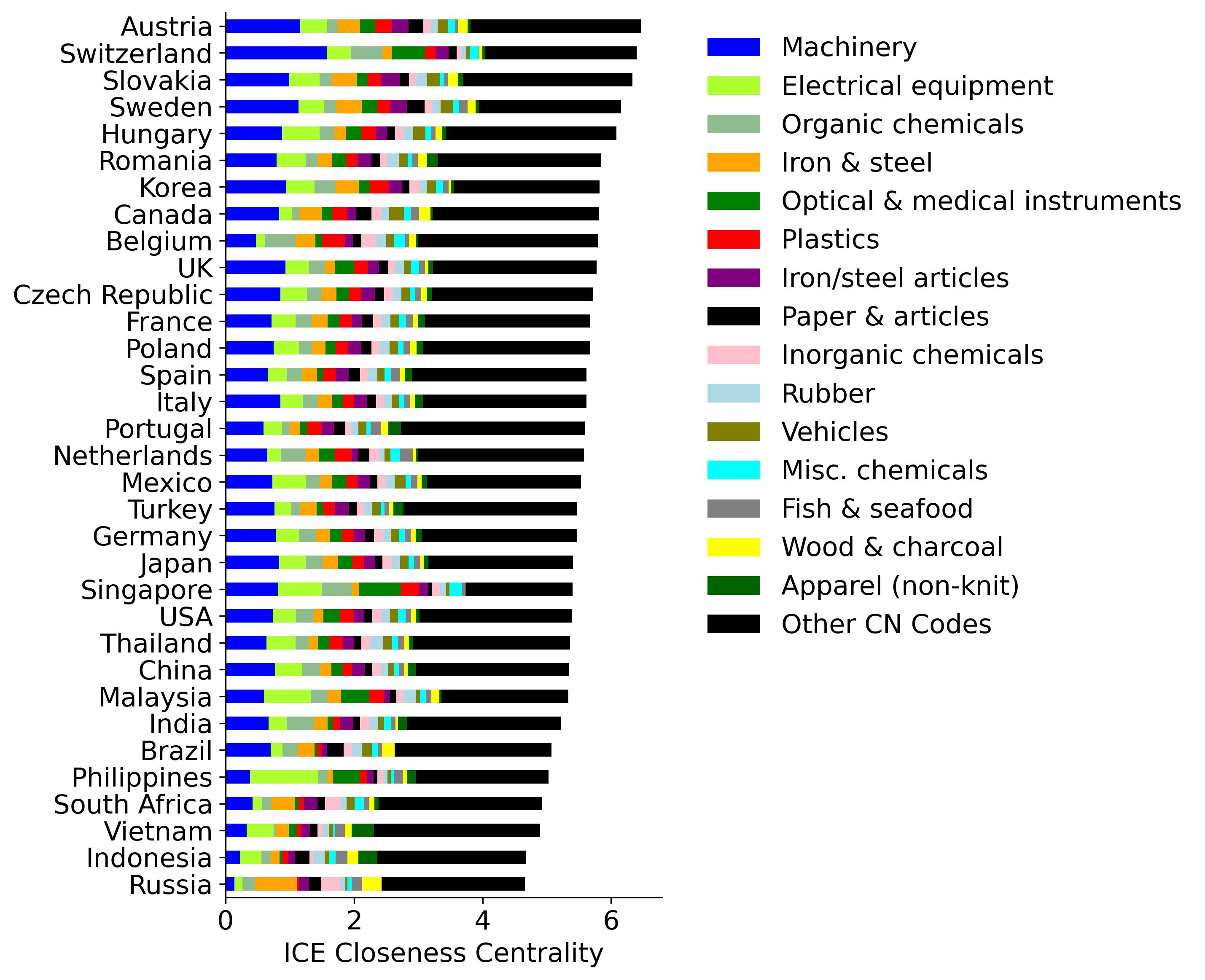}
    \caption{Closeness centrality of country specific products (RCA >= 1) to  ICE-specific products from the product spaced constructed from the industry-level product space. We show the shares per HS-codes for the 15 most influential HS-codes for countries with more than 150 firms that exist in both datasets.}
    \label{fig:closeness_ice_detail}
\end{figure}
\begin{figure}
    \centering
    \includegraphics[width=.99\linewidth]{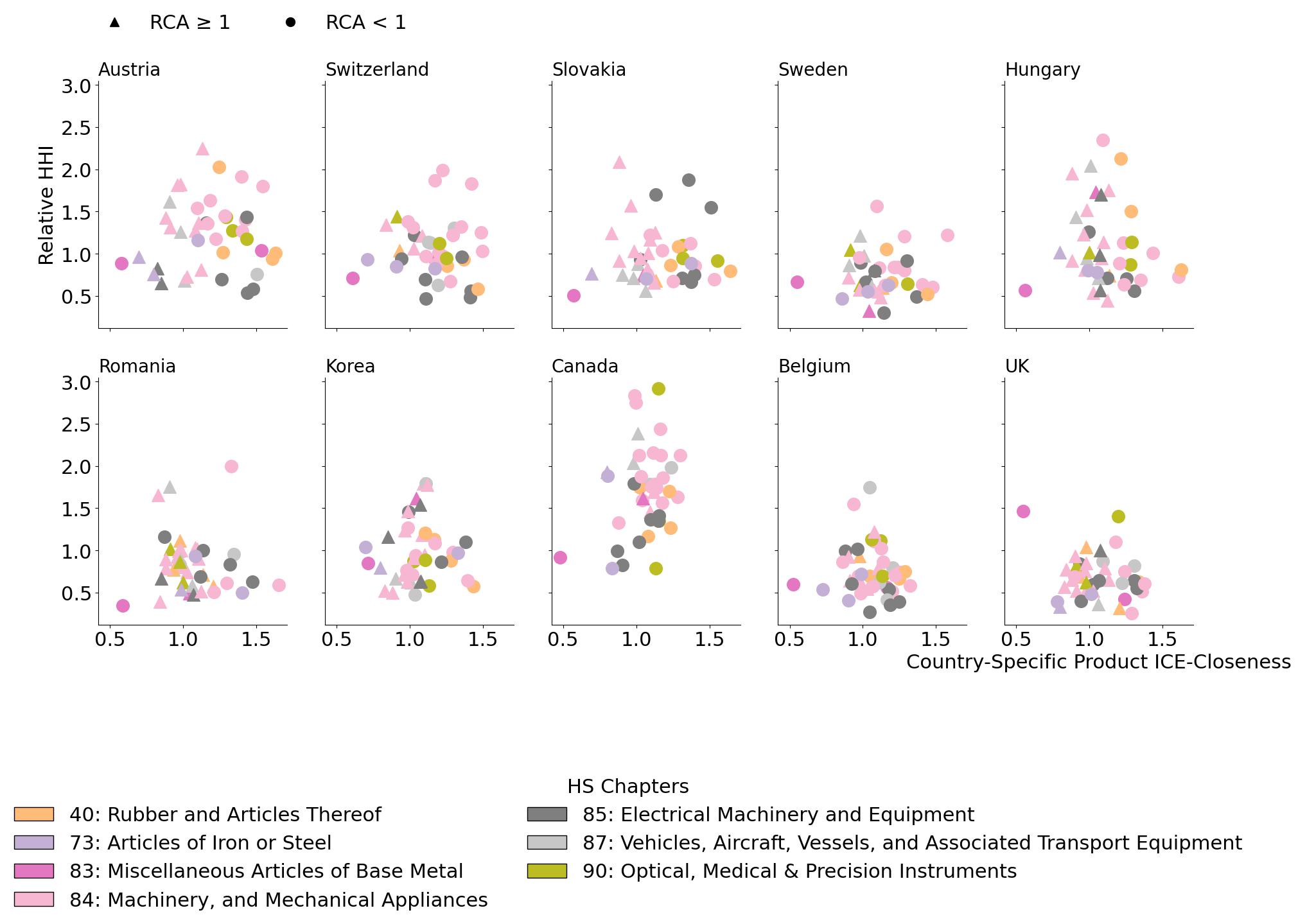}
    \caption{Herfindahl–Hirschman Index (HHI) and closeness of country specific products (RCA >= 1) with ICE specific HS codes from the product space constructed from the industry-level product space. Triangles indicate an RCA >= 1 and circles indicate an RCA < 1. }
    \label{fig:hhi_ice}
\end{figure}

\begin{figure}
    \centering
    \includegraphics[width=.99\linewidth]{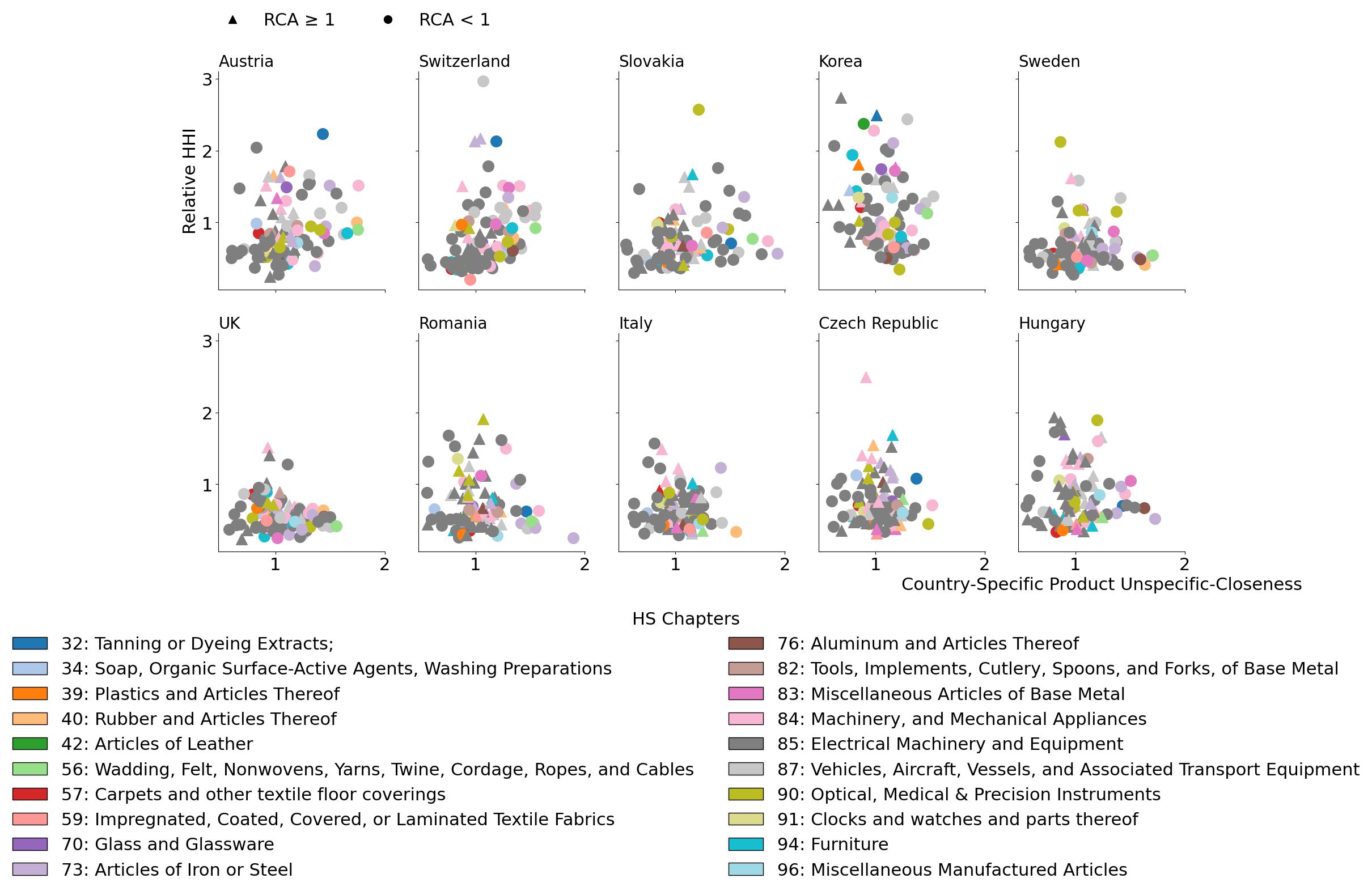}
    \caption{Herfindahl–Hirschman Index (HHI) and closeness of country specific products (RCA >= 1) with unspecific HS codes from the product space constructed from the industry-level product space. Triangles indicate an RCA >= 1 and circles indicate an RCA < 1. }
    \label{fig:hhi_unspec}
\end{figure}

\end{appendices}

\end{document}